\documentclass[11pt,fleqn]{article}
\usepackage{amsfonts, epsfig, amssymb, amsmath}
\usepackage[normalem]{ulem}
\usepackage{color}
\newcommand{\ol}{\setlength{\itemsep}{0pt.}\begin{enumerate}}
\newcommand{\eol}{\end{enumerate}\setlength{\itemsep}{-\parsep}}
\newcommand{\ignore}[1]{}
\title{{Hypercontractive inequalities for the second norm of highly concentrated functions, and Mrs. Gerber's-type inequalities for the second Renyi entropy}
\author{Niv Levhari, Alex Samorodnitsky}}
\begin{document}
\date{}
\maketitle

%  THEOREM-LIKE ENVIRONMENTS

\newtheorem{THEOREM}{Theorem}[section]
\newenvironment{theorem}{\begin{THEOREM} \hspace{-.85em} {\bf :}
}%
                        {\end{THEOREM}}
\newtheorem{LEMMA}[THEOREM]{Lemma}
\newenvironment{lemma}{\begin{LEMMA} \hspace{-.85em} {\bf :} }%
                      {\end{LEMMA}}
\newtheorem{COROLLARY}[THEOREM]{Corollary}
\newenvironment{corollary}{\begin{COROLLARY} \hspace{-.85em} {\bf
:} }%
                          {\end{COROLLARY}}
\newtheorem{PROPOSITION}[THEOREM]{Proposition}
\newenvironment{proposition}{\begin{PROPOSITION} \hspace{-.85em}
{\bf :} }%
                            {\end{PROPOSITION}}
\newtheorem{DEFINITION}[THEOREM]{Definition}
\newenvironment{definition}{\begin{DEFINITION} \hspace{-.85em} {\bf
:} \rm}%
                            {\end{DEFINITION}}
\newtheorem{EXAMPLE}[THEOREM]{Example}
\newenvironment{example}{\begin{EXAMPLE} \hspace{-.85em} {\bf :}
\rm}%
                            {\end{EXAMPLE}}
\newtheorem{CONJECTURE}[THEOREM]{Conjecture}
\newenvironment{conjecture}{\begin{CONJECTURE} \hspace{-.85em}
{\bf :} \rm}%
                            {\end{CONJECTURE}}
\newtheorem{MAINCONJECTURE}[THEOREM]{Main Conjecture}
\newenvironment{mainconjecture}{\begin{MAINCONJECTURE} \hspace{-.85em}
{\bf :} \rm}%
                            {\end{MAINCONJECTURE}}
\newtheorem{PROBLEM}[THEOREM]{Problem}
\newenvironment{problem}{\begin{PROBLEM} \hspace{-.85em} {\bf :}
\rm}%
                            {\end{PROBLEM}}
\newtheorem{QUESTION}[THEOREM]{Question}
\newenvironment{question}{\begin{QUESTION} \hspace{-.85em} {\bf :}
\rm}%
                            {\end{QUESTION}}
\newtheorem{REMARK}[THEOREM]{Remark}
\newenvironment{remark}{\begin{REMARK} \hspace{-.85em} {\bf :}
\rm}%
                            {\end{REMARK}}
%\newenvironment{proof}{\noindent {\bf Proof:} \hspace{.677em}}%
%                      {}

%theorem
\newcommand{\thm}{\begin{theorem}}
%lemma
\newcommand{\lem}{\begin{lemma}}
%proposition
\newcommand{\pro}{\begin{proposition}}
%definition
\newcommand{\dfn}{\begin{definition}}
%remark
\newcommand{\rem}{\begin{remark}}
%example
\newcommand{\xam}{\begin{example}}
%conjecture
\newcommand{\cnj}{\begin{conjecture}}
%main_conjecture
\newcommand{\mcnj}{\begin{mainconjecture}}
%problem
\newcommand{\prb}{\begin{problem}}
%question
\newcommand{\que}{\begin{question}}
%corollary
\newcommand{\cor}{\begin{corollary}}
%proof
\newcommand{\prf}{\noindent{\bf Proof:} }
%end theorem
\newcommand{\ethm}{\end{theorem}}
%end lemma
\newcommand{\elem}{\end{lemma}}
%end proposition
\newcommand{\epro}{\end{proposition}}
%end definition
\newcommand{\edfn}{\bbox\end{definition}}
%end remark
\newcommand{\erem}{\bbox\end{remark}}
%end example
\newcommand{\exam}{\bbox\end{example}}
%end conjecture
\newcommand{\ecnj}{\bbox\end{conjecture}}
%end main_conjecture
\newcommand{\emcnj}{\bbox\end{mainconjecture}}
%end problem
\newcommand{\eprb}{\bbox\end{problem}}
%end question
\newcommand{\eque}{\bbox\end{question}}
%end corollary
\newcommand{\ecor}{\end{corollary}}
%end proof
\newcommand{\eprf}{\bbox}
%begin equation
\newcommand{\beqn}{\begin{equation}}
%end equation
\newcommand{\eeqn}{\end{equation}}
% white box
\newcommand{\wbox}{\mbox{$\sqcap$\llap{$\sqcup$}}}
%black box
\newcommand{\bbox}{\vrule height7pt width4pt depth1pt}
\newcommand{\qed}{\bbox}
% \sup will be used for superscript.
\def\sup{^}

\def\H{\{0,1\}^n}

\def\S{S(n,w)}

\def\g{g_{\ast}}
\def\xop{x_{\ast}}
\def\y{y_{\ast}}
\def\z{z_{\ast}}

\def\f{\tilde f}

\def\n{\lfloor \frac n2 \rfloor}

\def \E{\mathop{{}\mathbb E}}
\def \R{\mathbb R}
\def \Z{\mathbb Z}
\def \F{\mathbb F}
\def \T{\mathbb T}

\def \x{\textcolor{red}{x}}
\def \r{\textcolor{red}{r}}
\def \Rc{\textcolor{red}{R}}

\def \noi{{\noindent}}

\def \iff{~~~~\Leftrightarrow~~~~}

\def \queq {\quad = \quad}

\def\<{\left<}
\def\>{\right>}
\def \({\left(}
\def \){\right)}

\def \e{\epsilon}
\def \l{\lambda}

\def\myblt{\noi --\, }

% Defining Tchebyshef polynomial
\def\Tp{Tchebyshef polynomial}
\def\Tps{TchebysDeto be the maximafine $A(n,d)$ l size of a code with distance $d$hef polynomials}
%right arrow
\newcommand{\rarrow}{\rightarrow}
%left arrow

\newcommand{\larrow}{\leftarrow}
%right arrow

\overfullrule=0pt
\def\setof#1{\lbrace #1 \rbrace}

\begin{abstract}

\noi Let $T_{\e}$, $0 \le \e \le 1/2$, be the noise operator acting on functions on the boolean cube $\H$. Let $f$ be a distribution on $\H$ and let $q > 1$. We prove tight Mrs. Gerber-type results for the second Renyi entropy of $T_{\e} f$ which take into account the value of the $q^{th}$ Renyi entropy of $f$. For a general function $f$ on $\H$ we prove tight hypercontractive inequalities for the $\ell_2$ norm of $T_{\e} f$ which take into account the ratio between $\ell_q$ and $\ell_1$ norms of $f$.

\end{abstract}

\section{Introduction}

\noi This paper considers the problem of quantifying the decrease in the $\ell_2$ norm of a function on the boolean cube when this function is acted on by the noise operator.

\noi Given a noise parameter $0 \le \e \le 1/2$, the noise operator $T_{\e}$ acts on functions on the boolean cube as follows: for $f:~\H \rarrow \R$, $T_{\e} f$ at a point $x$ is the expected value of $f$ at $y$, where $y$ is a random binary vector whose $i^{\small{th}}$ coordinate is $x_i$ with probability $1-\e$ and $1 - x_i$ with probability $\e$, independently for different coordinates. Namely, $\(T_{\e} f\)(x) =  \sum_{y \in \H} \e^{|y - x|}  (1-\e)^{n - |y-x|}  f(y)$, where $|\cdot|$ denotes the Hamming distance. We will write $f_{\e}$ for $T_{\e} f$, for brevity.

\noi Note that $f_{\e}$ is a convex combination of shifted copies of $f$. Hence, the noise operator decreases norms. Recall that the $\ell_q$ norm of a function is given by $\|f\|_q = \(\E |f|^q\)^{\frac 1q}$ (the expectations here and below are taken w.r.t. the uniform measure on $\H$). The norm sequence $\{\|f\|_q\}_q$ increases with $q$. An effective way to quantify the decrease of $\ell_q$ norm under noise is given by the hypercontractive inequality \cite{Bonami,Gross,Beckner} (see also e.g., \cite{O'Donnell} for background), which upperbounds the $\ell_q$ norm of the noisy version of a function by a smaller norm of the original function.
\beqn
\label{hypercontractive}
\|f_{\e}\|_q ~\le~ \|f\|_{1 + (1-2\e)^2(q-1)}.
\eeqn
This inequality is essentially tight in the following sense. For any $p < 1 + (q-1)(1-2\e)^2$ there exists a non-constant function $f: \H \rarrow \R$ with $\|f_{\e}\|_q >  \|f\|_p$.

\noi {\it Entropy} provides another example of a convex homogeneous functional on (nonnegative) functions on the boolean cube. For a nonnegative function $f$ let the entropy of $f$ be given by $Ent(f) = \E f \log_2 f -\E f \log_2 \E f$. The entropy of $f$ is closely related to Shannon's entropy of the corresponding distribution $f / \Sigma f$ on $\H$, and similarly the entropy of $f_{\e}$ is related to Shannon's entropy of the output of a binary symmetric channel with error probability $\e$ on input distributed according to $f / \Sigma f$ (see below and, e.g., the discussion in the introduction of \cite{RE}). The decrease in entropy (or, correspondingly, the increase in Shannon's entropy) after noise is quantified in the "Mrs. Gerber's Lemma" \cite{Wyner-Ziv}:
\beqn
\label{mrs gerber}
Ent\(f_{\e}\) ~\le~ n \E f \cdot \psi\(\frac{Ent(f)}{n \E f}, ~\e\),
\eeqn
where $\psi = \psi(x,\e)$ is an explicitly given function on $[0,1] \times \left[0,1/2\right]$, which is increasing and strictly concave in its first argument for any $0 < \e < \frac12$. This inequality is tight iff $f$ is a product function with equal marginals. That is, there exists a function $g: \{0,1\} \rarrow \R$, such that for any $x = \(x_1,...,x_n\) \in \H$ holds $f(x) = \prod_{i=1}^n g\(x_i\)$.

\noi One has $\psi(0,\e) = 0$ and $\frac{\partial \psi}{\partial x}_{|x=0} = (1-2\e)^2$. Hence $\psi(x,\e) \le (1-2\e)^2 \cdot x$, with equality only at $x = 0$. Hence the inequality (\ref{mrs gerber}) has the following weaker linear approximation version
\beqn
\label{mrs gerber - linear}
Ent\(f_{\e}\) ~\le~ (1-2\e)^2 \cdot Ent(f),
\eeqn
which is tight if and only if $f$ is a constant function.

\noi {\it R\'enyi entropies}. There is a well-known connection between $\ell_q$ norms of a nonnegative function $f$ and its entropy (see e.g., \cite{Cover-Thomas}): Assume, as we may by homogeneity, that $\E f = 1$. Then $Ent(f) = \lim_{q\rarrow 1} \frac{1}{q-1} \log_2 ||f||^q_q$. The quantity $Ent_q(f) = \frac{1}{q-1} \log_2 ||f||^q_q$ is known as the $q^{\small{th}}$ R\'enyi entropy of $f$ (\cite{Renyi}). The entropy sequence $\{Ent_q(f)\}_q$ increases with $q$. Restating the inequality (\ref{hypercontractive}) in terms of R\'enyi entropies gives
\[
Ent_q\(f_{\e}\) \le \frac{(1-2e)^2 q}{(1-2\e)^2(q-1)+1} \cdot Ent_{1 + (1-2\e)^2(q-1)}(f).
\]

\noi Note that taking $q \rarrow 1$ in this inequality recovers only the (weaker) linear approximation version (\ref{mrs gerber - linear}) of Mrs. Gerber's inequality (\ref{mrs gerber}). This highlights a important difference between inequalities (\ref{hypercontractive}) and (\ref{mrs gerber}). Mrs. Gerber's lemma takes into account the distribution of a function, specifically the ratio between its entropy and its $\ell_1$ norm. When this ratio is exponentially large in $n$, which typically holds in the information theory contexts in which this inequality is applied, (\ref{mrs gerber}) is significantly stronger than (\ref{mrs gerber - linear}). On the other hand, hypercontractive inequalities seem to be typically applied in contexts in which the ratio between different norms of the function is subexponential in $n$, and there are examples of such functions for which (\ref{hypercontractive}) is essentially tight. With that, there are several recent results \cite{PS, KS2, Y+} which show that (\ref{hypercontractive}) can be strengthened, if the ratio $\frac{\|f\|_q}{\|f\|_1}$, for some $q > 1$, is exponentially large in $n$. In the framework of R\'enyi entropies, the possibility of a result analogous to (\ref{mrs gerber}) for higher R\'enyi entropies was discussed in \cite{CT}.

\noi {\it Our results}. This paper proves a Mrs. Gerber type result for the second R\'enyi entropy, and a hypercontractive inequality for the $\ell_2$ norm of $f_{\e}$ which take into account the ratio between $\ell_q$ and $\ell_1$ norms of $f$. We try to pattern the results below after (\ref{mrs gerber}).

\noi We start with a Mrs. Gerber type inequality.

\pro
\label{pro:MrsG-2p}
Let $q > 1$, and let $f$ be a nonnegative function on $\H$ such that $\E f = 1$. Then
\beqn
\label{ineq:MrsG-2p}
\frac{Ent_2\(f_{\e}\)}{n} ~\le~ \psi_{2,q}\(\frac{Ent_q(f)}{n}, ~\e\),
\eeqn
where $\psi_{2,q}$ is an explicitly given function on $[0,1] \times \left[0,1/2\right]$, which is increasing and concave in its first argument.

\noi This inequality is essentially tight in the following sense. For any $0 < x < 1$ and $0 < \e < \frac12$, and for any $y < \psi_{2,q}(x,\e)$ there exists a sufficiently large $n$ and a nonnegative function $f$ on $\H$ with $\E f = 1$, $\frac{Ent_q(f)}{n} \le x$ and $\frac{Ent_2\(f_{\e}\)}{n} > y$.
\epro

\noi Let us make two comments about this result.

\myblt The functions $\{\psi_{2,q}\}_q$ are somewhat cumbersome to describe. Their precise definition will be given below.

\myblt Inequality (\ref{ineq:MrsG-2p}) upper bounds $Ent_2\(f_{\e}\)$ in terms of $Ent_q(f)$ for $q  > 1$, and $\e$. Taking $q = 2$ gives an upper bound on $Ent_2\(f_{\e}\)$ in terms of $Ent_2(f)$ and $\e$, in analogy to (\ref{mrs gerber}).

\myblt Recall that for a point $x \in \H$ and $0 \le r \le n$, the Hamming sphere of radius $r$ around $x$ is the set $\{y \in \H:~|y-x| = r\}$. As will be seen from the proof of Proposition~\ref{pro:MrsG-2p}, (\ref{ineq:MrsG-2p}) is essentially tight for a certain convex combination of the uniform distribution on $\H$ and the characteristic function of a Hamming sphere of an appropriate radius (depending on $q$, $\e$, and the required value of $Ent_q(f)$).

\myblt In information theory one typically considers a slightly different notion of R\'enyi entropies: For a probability distribution $P$ on $\Omega$, the $q^{\small{th}}$ Renyi entropy of $P$ is given by $H_q(P) = -\frac{1}{q-1} \log_2\(\sum_{\omega \in \Omega} P^q(\omega)\)$. To connect notions, if $f$ is a nonnegative (non-zero) function on $\H$ with expectation $1$, then $P = \frac{f}{2^n}$ is a probability distribution, and $Ent_q(f) = n - H_q(P)$. Furthermore, $Ent_q\(f_{\e}\) = n - H_q\(X \oplus Z\)$, where $X$ is a random variable on $\H$ distributed accordinng to $P$ and $Z$ is an independent noise vector corresponding to a binary symmetric channel with crossover probability $\e$. Hence, (\ref{mrs gerber}) can be restated as
\[
H\(X \oplus Z\) \ge n \cdot \varphi\(\frac{H(X)}{n}, ~\e\),
\]
and Proposition~\ref{pro:MrsG-2p} can be restated as
\[
H_2\(X \oplus Z\) \ge n \cdot \varphi_{2,q}\(\frac{H_q(X)}{n}, ~\e\)
\]
Here $\varphi$ is an explicitly given function on $[0,1] \times \left[0,1/2\right]$, which is increasing and convex in its first argument ($\varphi(x,\e)  = 1 - \psi(1-x,\e)$), and similarly for $\varphi_{2,q}$.

\noi Next, we describe our main result, a hypercontractive inequality for the $\ell_2$ norm of $f_{\e}$ which takes into account the ratio between $\ell_q$ and $\ell_1$ norms of $f$, and more specifically $Ent_q\(\frac{f}{\|f\|_1}\) = \frac{q}{q-1} \log_2\(\frac{\|f\|_q}{\|f\|_1}\)$.

\thm
\label{thm:NHC}
Let $q > 1$, and let $f$ be a non-zero function on $\H$. Then
\beqn
\label{ineq:NHC}
\|f_{\e}\|_2 ~\le~ \|f\|_{\kappa},
\eeqn
where $\kappa = \kappa_{2,q}\(\frac{Ent_q\(\frac{f}{\|f\|_1}\)}{n},\e\)$, and $\kappa_{2,q}$ is an explicitly given function on $[0,1] \times \left[0,1/2\right]$, which is decreasing in its first argument and which satisfies $\kappa_{2,q}(0,\e) = 1 + (1-2\e)^2$, for all $0 \le \e \le \frac12$.

\noi This inequality is essentially tight in the following sense. For any $0 < x < 1$ and $0 < \e < \frac12$, and for any $y < \kappa_{2,q}(x,\e)$ there exists a sufficiently large $n$ and a function $f$ on $\H$ with $\frac{Ent_q(f)}{n} \ge x$ and $\|f_{\e}\|_2 > \|f\|_y$.
\ethm

\noi Some comments (see also Lemma~\ref{lem:comm-thm} below).

\myblt The precise definition of the functions $\{\kappa_{2,q}\}_q$ will be given below. At this point let us just observe that since the sequence $\{Ent_q(f)\}_q$ increases with $q$, we would expect the fact that $Ent_q(f)$ is large to become less significant as $q$ increases. This is expressed in the properties of the functions $\{\kappa_{2,q}\}_q$ in the following manner: If $q \ge 2$ then for any $0 < \e < \frac12$ the function $\kappa_{2,q}(x,\e)$ starts as a constant-$\(1 + (1-2\e)^2\)$ function up to some $x = x(q,\e) > 0$, and becomes strictly decreasing after that. In other words $x(q,\e)$  is the largest possible value of $\frac{Ent_q\(\frac{f}{\|f\|_1}\)}{n}$ for which Theorem~\ref{thm:NHC} provides no new information compared to (\ref{hypercontractive}). For $1 < q < 2$ there is a value $0 < \e(q) < \frac12$, such that for all $\e \le \e(q)$ the function $\kappa_{2,q}(x,\e)$ is strictly decreasing (in which case we say that $x(q,\e) = 0$). However, $x(q,\e) >  0$ for all $\e > \e(q)$. The function $\e(q)$ decreases with $q$ (in particular, $\e(q) = 0$ for $g \ge 2$). The function $x(q,\e)$ increases both in $q$ and in $\e$.

\myblt Notably, taking $q \rarrow 1$ in Theorem~\ref{thm:NHC} gives (see Corollary~\ref{cor:useful})
\[
\|f_{\e}\|_2 ~\le~ \|f\|_{\kappa},
\]
where $\kappa = \kappa_{2,1}\(Ent\(\frac{f}{\|f\|_1}\)/n,\e\) = -\frac{Ent\(\frac{f}{\|f\|_1}\)/n}{\phi_{\e}\(1 - Ent\(\frac{f}{\|f\|_1}\)/n\)}$. The function $\kappa_{2,1}(x,\e) = -\frac{x}{\phi_{\e}(1-x)}$ is strictly decreasing in $x$ for any $0 < \e < \frac12$. It satisfies $\kappa_{2,1}(0,\e) = \lim_{x \rarrow 0} \kappa_{2,1}(x,\e) = 1 + (1-2\e)^2$, for all $0 \le \e \le \frac12$. Hence, this is stronger than (\ref{hypercontractive}) for any non-constant function $f$ and for any $0 < \e < \frac12$, with the difference between the two inequalities becoming significant when $Ent\(\frac{f}{\|f\|_1}\)/n$ is bounded away from $0$.

\myblt As will be seen from the proof of Theorem~\ref{thm:NHC}, (\ref{ineq:NHC}) is essentially tight for a certain convex combination of the uniform distribution on $\H$ and characteristic functions of one or two Hamming spheres of appropriate radii (the number of the spheres and their radii depend on $q$, $\e$, and the required value of $Ent_q\(\frac{f}{\|f\|_1}\)$).

\myblt Let $f$ be a non-constant function and let $0 < \e < \frac12$ be fixed. Consider the function $F(q) = F_{f,\e}(q) = \kappa_{2,q}\(\frac{Ent_q\(\frac{f}{\|f\|_1}\)}{n},\e\)$. It will be seen that there is a unique value $1 < q(f,\e) \le 1 + (1-2\e)^2$ of $q$ for which $F(q) = q$. Furthermore, $q(f,\e) = \min_{q \ge 1} F(q)$. Hence it provides the best possible value for $\kappa$ in Theorem~\ref{thm:NHC}. With that, determining $q(f,\e)$ might in principle require knowledge of all the Renyi entropies $Ent_q(f)$, for $1 \le q \le 1 + (1-2\e)^2$, while typically we are in possession of one of the "easier" Renyi entropies, such as $Ent(f)$ or $Ent_2(f)$.

\noi {\it Logarithmic Sobolev inequalities}. Viewing both sides of (\ref{hypercontractive}) as functions of $\e$, and writing $L(\e)$ for the LHS and $R(\e)$ for the RHS, we have $L(0) = R(0) = \|f\|_2$, and $L(\e) \le R(\e)$ for $0 \le \e \le \frac12$. Since both $L$ and $R$ are differentiable in $\e$ this implies $L'(0) \le R'(0)$. This inequality is the logarithmic Sobolev inequality for the Hamming cube \cite{Gross}. We proceed to describe it in more detail. Recall that the Dirichlet form ${\cal E}(f,g)$ for functions $f$ and $g$ on the Hamming cube is defined by ${\cal E}(f,g) = \E_x \sum_{y \sim x} \Big(f(x) - f(y)\Big) \Big(g(x) - g(y)\Big)$. Here $y \sim x$ means that $x$ and $y$ differ in precisely one coordinate. The logarithmic Sobolev inequality then states that ${\cal E}(f,f) \ge 2 \ln 2 \cdot Ent\(f^2\)$. Applying the same approach to the inequalities of Theorem~\ref{thm:NHC} leads to a family of logarithmic Sobolev inequalities of the form ${\cal E}(f,f) \ge c \cdot Ent\(f^2\)$, where the constant $c$ depends on $Ent_q(f)$ and belongs to the interval $[2 \ln 2, 2]$. Going back to the preceding remark, it is not hard to see that for any function $f$ holds $q(f,\e) \rarrow_{\e \rarrow 0} 2$, and hence we take $q = 2$ in Theorem~\ref{thm:NHC} to obtain the following claim. Here and below we write $H(t) = t \log_2\(\frac 1t\) + (1-t) \log_2\(\frac{1}{1-t}\)$ for the binary entropy function.

\cor
\label{cor:Lsob}
For any function $f$ on $\H$ holds
\[
{\cal E}(f,f) ~\ge~ C\(\frac{Ent_2\(\frac{f}{\|f\|_1}\)}{n}\) \cdot Ent\(f^2\),
\]
where $C(x) = 2 \cdot \frac{1 - 2\sqrt{H^{-1}(1-x) \(1 - H^{-1}(1-x)\)}}{x}$ is a convex and increasing function on $[0,1]$, taking $[0,1]$ onto $\left[2 \ln 2, 2\right]$.
\ecor

\noi The function $C$ was defined in \cite{Modified Log-Sobolev}, where a somewhat stronger logarithmic Sobolev inequality ${\cal E}(f,f) \ge C\(\frac{Ent\(\frac{f^2}{\|f\|^2_2}\)}{n}\) \cdot Ent\(f^2\)$ was shown, using a different approach.\footnote{It seems that it might be possible to recover the inequality of \cite{Modified Log-Sobolev} by differentiating a corresponding hypercontractive inequality at zero, if one considers a more general version of Theorem~\ref{thm:NHC} which takes into account the ratio between $\ell_q$ and $\ell_p$ norms of $f$, for $q > p$ (and in this case taking both $q$ and $p$ to be very close to $2$). We omit the details.}
 This was used in \cite{Modified Log-Sobolev} to establish the following claim, answering a question of \cite{FT}. Let $A \subseteq \H$. Let $M_A$ be the adjacency matrix of the subgraph of the discrete cube induced by the vertices of $A$. Then the maximal eigenvalue $\l(A)$ of $M_A$ satisfies
\beqn
\label{ineq:eigen}
\l(A) ~\le~ C\(\frac 1n \log_2\(\frac{2^n}{|A|}\)\) \cdot \log_2\(\frac{2^n}{|A|}\).
\eeqn
This is almost tight if $A$ is a Hamming ball of exponentially small cardinality. We observe that (\ref{ineq:eigen}) is also an easy implication of Corollary~\ref{cor:Lsob} (following the argument of  \cite{Modified Log-Sobolev}), and hence it might be viewed as another consequence of Theorem~\ref{thm:NHC}.

\noi {\it Full statements of Proposition~\ref{pro:MrsG-2p} and Theorem~\ref{thm:NHC}}

\noi We now define the functions $\{\psi_{2,q}\}_q$ in Proposition~\ref{pro:MrsG-2p} and $\{\kappa_{2,q}\}_q$ in Theorem~\ref{thm:NHC}, completing the statements of these claims. We start with introducing yet another function on $[0,1] \times \left[0,1/2\right]$ which will play a key role in what follows (we remark that this function was studied in \cite{KS2}). For $0 \le x \le 1$ and $0 \le \e \le \frac12$, let $\sigma = H^{-1}(x)$ and let $y = y(x, \e) = \frac{-\e^2 + \e \sqrt{\e^2 + 4(1-2\e) \sigma (1-\sigma)}}{2(1-2\e)}$. Let
\[
\Phi(x, \e) ~=~ \frac12 \cdot \(x - 1 + \sigma H\(\frac{y}{\sigma}\) + (1-\sigma) H\(\frac{y}{1-\sigma}\) + 2y\log_2(\e) + (1-2y) \log_2(1-\e)\).
\]

\noi The function $\Phi$ is nonpositive. It is increasing and concave in its first argument. Additional relevant properties of $\Phi$ are listed in Lemma~\ref{lem:phi-technical} below. For a fixed $\e$, it will be convenient to write $\phi_{\e}(x) = \Phi\big(x,2\e(1-\e)\big)$, viewing $\phi_{\e}$ as a univariate function on $[0,1]$.

\dfn
\label{dfn:psi-kappa}

\noi Let $0 \le x \le 1$ and $0 \le \e \le \frac12$.

\begin{itemize}

\item If $\phi_{\e}'(1-x) < \frac 1q$, let $\alpha_0 = \(\phi'_{\e}\)^{-1}\(\frac 1q\)$. Define
\[
\psi_{2,q}(x,\e) ~=~ 2 \cdot \left\{\begin{array}{ccc} \frac{q-1}{q} \cdot x + \(\phi_{\e}\(\alpha_0\) + \frac{1 - \alpha_0}{q}\) & \mbox{if} & \phi_{\e}'(1-x) < \frac 1q \\ \phi_{\e}(1 - x) + x & \mbox{otherwise} \end{array}\right.
\]

\item Let $y = \frac{q-1}{q} \cdot x + \frac 1q$. Let $q_0 = 1 + (1-2\e)^2$. If $y \ge \frac{1}{q_0}$, let $\alpha_0$ be determined by $1 - \alpha_0 - \frac{\alpha_0 \phi_{\e}(\alpha_0)}{1-\alpha_0} = y$. If $x = 0$, define $\kappa_{2,q}(x,\e) = q_0$. Otherwise, define

\[
\kappa_{2,q}(x,\e) ~=~ \left\{\begin{array}{ccc} q_0 & \mbox{if} & y \le \frac{1}{q_0} \\ -\frac{x}{\phi_{\e}(1 - x)} & \mbox{if} & y >  \frac{1}{q_0} \mbox{~and~} -\frac{x}{\phi_{\e}(1 - x)} \ge q \\
\frac{\alpha_0 - 1}{\phi_{\e}(\alpha_0)} & \mbox{if} & y >  \frac{1}{q_0} \mbox{~and~} -\frac{x}{\phi_{\e}(1 - x)} < q \end{array}\right.
\]

\end{itemize}

\edfn

\noi We remark that it is not immediately obvious that the functions $\psi_{2,q}$ and $\kappa_{2,q}$ are well-defined. This will be clarified in the proofs of Proposition~\ref{pro:MrsG-2p} and Theorem~\ref{thm:NHC}.

\noi We state explicitly two special cases of Theorem~\ref{thm:NHC}, which seem to be the most relevant for applications. They describe the improvement over (\ref{hypercontractive}), given non-trivial information about $Ent(f)$ and $\|f\|_2$.

\cor
\label{cor:useful}
\begin{enumerate}

\item Taking $q \rarrow 1$ in Theorem~\ref{thm:NHC} gives:
\[
\|f_{\e}\|_2 ~\le~ \|f\|_{\kappa}, \quad \mbox{with} \quad \kappa = -\frac{Ent\(\frac{f}{\|f\|_1}\)/n}{\phi_{\e}\(1 - Ent\(\frac{f}{\|f\|_1}\)/n\)}.
\]

\item  Taking $q = 2$ in Theorem~\ref{thm:NHC} gives, for $x = \frac{Ent_2\(\frac{f}{\|f\|_1}\)}{n}$ and $q_0 = 1 + (1-2\e)^2$
\[
\|f_{\e}\|_2 ~\le~ \|f\|_{\kappa}, \quad \mbox{with} \quad \kappa = \left\{\begin{array}{ccc} q_0 & \mbox{if} & \frac{x+1}{2} \le \frac{1}{q_0} \\ \frac{\alpha - 1}{\phi_{\e}(\alpha)} & \mbox{otherwise} \end{array}\right.
\]
In the second case $\alpha$ is determined by $1 - \alpha - \frac{\alpha \phi_{\e}(\alpha)}{1-\alpha} = \frac{x+1}{2}$.
\end{enumerate}
\ecor

\noi We observe that both Proposition~\ref{pro:MrsG-2p} and Theorem~\ref{thm:NHC} are based on the following claim (\cite{KS2}, Corollary~3.2;  this claim also explains the relevance of function $\Phi$).

\thm
\label{thm:bounded-support}
\noi Let $0 \le x \le 1$. Let $f$ be a function on $\H$ supported on a set of cardinality at most $2^{xn}$. Then, for any $0 \le \e \le \frac12$ holds
\[
\<f_{\e} ,f\> ~\le~ 2^{\(2\Phi\(x,\e\) + 1 - x\) \cdot n} \cdot \|f\|^2_2,
\]
Moreover, this is tight, up to a polynomial in $n$ factor, if $f$ is the characteristic function of a Hamming sphere or radius $H^{-1}(x) \cdot n$.
\ethm

\noi {\it Related work}

\noi In \cite{PS} it was shown that if $\frac{\|f\|_p}{\|f\|_1} \ge 2^{\rho n}$, for some $p \ge 1$ and $\rho \ge 0$, then
$\|f\|_p \ge \|f_{\e}\|_{1 + \frac{p-1}{(1-2\e)^2} + \Delta(p,\rho,\e)}$,
where $\Delta(p,\rho,\e) > 0$ for all $p > 1$, $\e, \rho > 0$ (cf. with (\ref{hypercontractive}), which can be restated as $\|f\|_p \ge \|f_{\e}\|_{1 + \frac{p-1}{(1-2\e)^2}}$, for $p = 1 +  (1-2\e)^2 (q-1)$). The function $\Delta(p,\rho,\e)$ is "semi-explicit", in the following sense: it is an explicit function of the (unique) solution of a certain explicit differential equation.

\noi In \cite{Y+} it was shown, using a different approach, that (restating the result in the notation of this paper) $\|f_{\e}\|_2 \le \|f\|_q$, where $q$ is determined by $F_{f,\e}(q) =  q$ (in the notation of the last comment above to Theorem~\ref{thm:NHC}). As we have observed, this is the best possible value for $\kappa$ in Theorem~\ref{thm:NHC}, but it might not be easy to determine explicitly in practice (compare with Corollary~\ref{cor:useful}).

\noi This paper is organized as follows. We prove Proposition~\ref{pro:MrsG-2p} in Section~\ref{sec:pro} and Theorem~\ref{thm:NHC} in Section~\ref{sec:thm}. We prove the remaining claims, including some technical lemmas and claims made above in the comments to the main results, in Section~\ref{sec:remaining}.

\section{Proof of Proposition~\ref{pro:MrsG-2p}}
\label{sec:pro}

\noi We first prove (\ref{ineq:MrsG-2p}) and then show it to be tight. We prove (\ref{ineq:MrsG-2p}) in two steps, using Theorem~\ref{thm:bounded-support} to reduce it to a claim about properties of the function $\phi_{\e}$, and then proving that claim.

\noi We start with the first step. It follows closely the proof of Theorem~1.8 in \cite{KS2}, and hence will be presented rather briefly, and not in a self-contained manner. Let $f$ be a function on $\H$, for which we want to show (\ref{ineq:MrsG-2p}). We may assume, w.l.o.g., that $f$ is positive, in fact that $f \ge 2^{-n}$. Furthermore, Theorem~\ref{thm:bounded-support} implies the following fact. We can partition $\H$ into $O(n)$ level sets $A_1,...,A_r$ of $f$, setting $\alpha_i = \frac 1n \log_2\(|A_i|\)$, and $\nu_i = \frac 1n \log_2\(v_i\)$, where $v_i$ is the minimal value of $f$ on $A_i$, so that, up to a negligible (vanishing with $n \rarrow \infty$) error holds.
\[
\frac{Ent_2\(f_{\e}\)}{n} ~=~ \frac 1n \log_2 \|f_{\e}\|^2_2 ~\le~ 2 \cdot \max_{1 \le i \le r} \Big\{\phi_{\e}\(\alpha_i\) + \nu_i\Big\}.
\]
The negligible error we have mentioned contributes towards a negligible error in (\ref{ineq:MrsG-2p}), which can then be removed by a tensorization argument, so we will ignore it from now on.

\noi Let $N = \frac 1n \log_2\(\|f\|_q\)$. Note that $N = \frac{q-1}{q} \cdot \frac{Ent_q(f)}{n}$. Hence, in particular, $N \le \frac{q-1}{q}$. Note also that for any $1 \le i \le r$ holds $\alpha_i  + \nu_i \le 1$ (since $\E f = 1$) and  $\frac{\alpha_i - 1}{q} + \nu_i \le N$. We also have $0 \le \alpha_i \le 1$ and $-1 \le \nu_i \le 1$. This discussion leads to the definition of the following two subsets of $\R^2$, which will play an important role in the proof of Theorem~\ref{thm:NHC} as well. (We remark that the relevance of the set $\Omega$ in the following definition is not immediately obvious. It will be made clear in the following arguments.)

\dfn
\label{dfn:Omega}
Let $q > 1$ and $0 < N \le \frac{q-1}{q}$. Let $\Omega_0 \subseteq \R^2$ be defined by
\[
\Omega_0 ~=~ \left\{(\alpha, \nu):~0 \le \alpha \le 1, ~ -1 \le \nu \le 1, ~\alpha + \nu \le 1, ~\frac{\alpha-1}{q} + \nu \le N \right\}.
\]

\noi Let $\Omega \subseteq \Omega_0$ be the set of all pairs $(\alpha, \nu) \in \Omega_0$ with $\nu \ge 0$.
\edfn

\noi By the preceding discussion, (\ref{ineq:MrsG-2p}) will follow from the following claim.

\lem
\label{lem:MrsG-2p-phi}
For all $0 \le \e \le \frac12$ holds
\[
\max_{(\alpha,\nu) \in \Omega_0} \Big\{\phi_{\e}(\alpha) + \nu\Big\} ~=~ \frac12 \cdot \psi_{2,q}\(\frac{qN}{q-1},\e\),
\]
where $\psi_{2,q}$ is defined in Definition~\ref{dfn:psi-kappa}.
\elem

\noi Before proving Lemma~\ref{lem:MrsG-2p-phi}, we collect the relevant properties of the function $\phi_{\e}$ in the following lemma.

\lem
\label{lem:phi-technical}

\noi Let $0 < \e < \frac12$. Let $q_0 = q_0(\e) = 1 + (1-2\e)^2$. The function $\phi_{\e}$ has the following properties.

\begin{enumerate}

\item $\phi_{\e}(\alpha)$ is strictly concave and increasing from $\phi_{\e}(0)=  -\frac{\log_2\(\frac{4}{q_0}\)}{2}$ to $0$ on $0 \le \alpha \le 1$.

\item $\phi_{\e}'(0) = 1$, $\phi_{\e}'(1) = \frac{1}{q_0}$.

\item $\frac{\alpha - 1}{\phi_{\e}(\alpha)}$ is strictly increasing in $\alpha$, going up to $q_0$, as $\alpha \rarrow 1$.

\item The function $g(\alpha) = 1 - \alpha - \frac{\alpha}{1-\alpha} \cdot \phi_{\e}\(\alpha\)$ is strictly decreasing on $[0,1]$. Moreover, $g(0) = 1$ and $g(1) = \frac{1}{q_0}$.

\end{enumerate}

\elem

\noi This lemma will be proved below. For now we assume its correctness, and proceed with the proof of Lemma~\ref{lem:MrsG-2p-phi}

\prf

\noi Our first observation is that the maximum of $\phi_{\e}(\alpha) + \nu$ on $\Omega_0$ is located in $\Omega$, since for any point $(\alpha, \nu) \in \Omega_0$ with $\nu < 0$, the point $(\alpha, 0)$ is in $\Omega$. So we may and will replace $\Omega_0$ with $\Omega$ in the following argument.

\noi Since $\phi_{\e}$ is increasing, any local maximum of $\phi_{\e}(\alpha) + \nu$ is located on the upper boundary of $\Omega$, that is on the piecewise linear curve which starts as the straight line $\frac{\alpha}{q} + \nu = N + \frac 1q$, for $0 \le \alpha \le 1 - \frac{qN}{q-1}$ and continues as the straight line $\alpha + \nu = 1$ for $1 - \frac{qN}{q-1} \le \alpha \le 1$.

\noi Note that, since $\phi_{\e}' < 1$ for $\alpha > 0$, the function $\phi_{\e}(\alpha) + \nu$ decreases (as a function of $\alpha$) on the line $\alpha + \nu = 1$ for $1 - \frac{qN}{q-1} \le \alpha \le 1$. Next, let $h(\alpha) = \phi_{\e}(\alpha) - \frac{\alpha}{q} + \(N + \frac 1q\)$. The function $h$ describes the restriction of $\phi_{\e}(\alpha) + \nu$ to the line $\frac{\alpha}{q} + \nu = N + \frac 1q$, and we are interested on the maximum of $h$ on the interval $I = \left\{0 \le \alpha \le 1 - \frac{qN}{q-1}\right\}$. We have $h'(\alpha) = \phi_{\e}'(\alpha) - \frac 1q$. By Lemma~\ref{lem:phi-technical}, the function $h$ is concave, and hence there are two possible cases:

\begin{itemize}

\item $\phi_{\e}'\(1 - \frac{qN}{q-1}\) \ge \frac 1q$. In this case $h$ is increasing on $I$ and we get
\[
\max_{(\alpha,\nu) \in \Omega} \Big\{\phi_{\e}(\alpha) + \nu\Big\} ~=~ \max_{\alpha \in I} \{h(\alpha)\} ~=~ h\(1 - \frac{qN}{q-1}\) ~=~
\]
\[
\phi_{\e}\(1 - \frac{qN}{q-1}\) + \frac{qN}{q-1} ~=~   \frac12 \cdot \psi_{2,q}\(\frac{qN}{q-1},\e\).
\]
The last equality follows from the definition of $\psi_{2,q}$ in this case.

\item $\phi_{\e}'\(1 - \frac{qN}{q-1}\) < \frac 1q$. Note that, by Lemma~\ref{lem:phi-technical}, $1 = \phi_{\e}'(0) > \frac 1q$. Hence, in this case the maximum of $h$ on $I$ is located at the unique zero of its derivative, that is at the point $\alpha_0$ such that $\phi_{\e}'\(\alpha_0\) = \frac 1q$. Using the definition of $\psi_{2,q}$ in this case, we get
\[
\max_{(\alpha,\nu) \in \Omega} \Big\{\phi_{\e}(\alpha) + \nu\Big\} ~=~ h\(\alpha_0\) ~=~ N + \(\phi_{\e}\(\alpha_0\) + \frac{1 - \alpha_0}{q}\) ~=~ \frac12 \cdot \psi_{2,q}\(\frac{qN}{q-1},\e\).
\]
\end{itemize}
\eprf

\noi This concludes the proof of (\ref{ineq:MrsG-2p}). The fact that $\psi_{2,q}\(x,\e\)$ is strictly increasing and concave in its first argument is an easy implication of Lemma~\ref{lem:phi-technical}.

\noi We pass to showing the tightness of (\ref{ineq:MrsG-2p}). Let $0 < \e < \frac12$ and $0 < x < 1$. Set $N = \frac{q-1}{q} \cdot x$. Let $\Omega$ be the domain defined in Definition~\ref{dfn:Omega}, and let $\(\alpha^{\ast}, \nu^{\ast}\)$ be the maximum point of $\phi_{\e}(\alpha) + \nu$ on $\Omega$ (note that the discussion above determines this point uniquely).  We proceed to define the function $f$. Let $n$ be sufficiently large. For $y \in \H$, let $|y|$ denotes the Hamming weight of $y$, that is the number of $1$-coordinates in $y$. Let $r = \lfloor H^{-1}\(\alpha^{\ast}\) \cdot n\rfloor$. Let $S = \{y \in \H, |y| = r\}$ be the Hamming sphere around zero of radius $r$ in $\H$. Now there are two cases to consider.

\begin{itemize}

\item If $\phi_{\e}'(1-x) < \frac 1q$, then by the discussion above, the point $\(\alpha^{\ast}, \nu^{\ast}\)$ lies on the line $\frac{\alpha}{q} + \nu = N + \frac 1q$, but not on the line $\alpha + \nu = 1$. Observe that $2^{\alpha^{\ast} n - o(n)} \le |S| \le 2^{\alpha^{\ast} n}$  (the first estimate follows from the Stirling formula, for the second estimate see e.g., Theorem~1.4.5. in \cite{van Lint}). As the first attempt, let $g = 2^{\nu^{\ast} n} \cdot 1_S$. Then $N - o(n) \le \frac{\alpha^{\ast}-1}{q} + \nu^{\ast} - o(n) \le \frac 1n \log_2 \|g\|_q \le \frac{\alpha^{\ast}-1}{q} + \nu^{\ast} = N$. That is,  $x - o_n(1) \le \frac{Ent_q(g)}{n} \le x$. However, $\E g$ is exponentially small. To correct that, we define $f$ to be $v = 2^{\(\nu^{\ast} - \delta\) \cdot n}$ on $S$, and $\frac{2^n - |S| v}{2^n - |S|}$ on the complement of $S$. Then $\E f = 1$. We choose $\delta$ to be as small as possible, while ensuring that $\frac{Ent_q(f)}{n} \le x$. Since the contribution of the constant-$1$ function to $\|f\|_q$ is exponentially small w.r.t. $\|f\|_q$, we can choose $\delta = o_n(1)$. We now have $\E f = 1$, $\frac{Ent_q(f)}{n} \le x$, and
    \[
    \frac{Ent_2(f_{\e})}{n} ~=~ \frac 1n \log_2 \|f_{\e}\|^2_2 ~=~ \frac 1n \log_2 \<f_{2\e(1-\e)}, f\> ~\ge~
    \]
    \[
    2 \cdot \(\phi_{\e}\(\alpha^{\ast}\) + \nu^{\ast}\) - o_n(1) ~\ge~ \psi_{2,q}(x,\e) - o_n(1).
    \]
 Here the second equality follows from the semigroup property of the noise operator: $T_{\e} \circ T_{\e} = T_{2\e(1-\e)}$. The first inequality follows from the tightness part of Theorem~1.6 and the definition of $\phi_{\e}$. The second inequality follows from Lemma~\ref{lem:MrsG-2p-phi}.

 \noi The tightness of (\ref{ineq:MrsG-2p}) in this case now follows, taking into account the fact that $\psi_{2,q}$ is strictly increasing.

 \item If $\phi_{\e}'(1-x) \ge \frac 1q$, the point $\(\alpha^{\ast}, \nu^{\ast}\)$ lies on the intersection of the lines $\frac{\alpha}{q} + \nu = N + \frac 1q$, and $\alpha + \nu = 1$. Hence the function $g = 2^{\nu^{\ast} n} \cdot 1_S$ has both $x - o_n(1) \le \frac{Ent_q(g)}{n} \le x$, and $1 - o_n(1) \le \E g \le 1$. It is easy to see that $g$ can be corrected as in the preceding case, by decreasing it slightly on $S$ and adding a constant component, to obtain a function $f$ with expectation $1$ and $Ent_q(f) \le x$, and with $ \frac{Ent_2(f_{\e})}{n} \ge \psi_{2,q}(x,\e) - o_n(1)$, proving the tightness of (\ref{ineq:MrsG-2p}) in this case as well. We omit the details.
\end{itemize}

\section{Proof of Theorem~\ref{thm:NHC}}
\label{sec:thm}

\noi The high-level outline of the argument in this proof is similar to that of Proposition~\ref{pro:MrsG-2p}. We start with proving (\ref{ineq:NHC}), doing this in two steps. In the first step Theorem~\ref{thm:bounded-support} is used to reduce (\ref{ineq:NHC}) to a claim about properties of the function $\phi_{\e}$. That claim is proved in the second step.

\noi We will give only a brief description of the first step since, similarly to the first step in the proof of Proposition~\ref{pro:MrsG-2p}, it follows closely the proof of Theorem~1.8 in \cite{KS2}. Let $f$ be a function on $\H$, for which we may and will assume that $f \ge 2^{-n}$ and that $\E f = \|f\|_1 = 1$. There are $O(n)$ real numbers $0 \le \alpha_1, ..., \alpha_r \le 1$ and $-1 \le \nu_1,...,\nu_r \le 1$, such that, up to a negligible error, which may be removed by tensorization, we have
\[
\frac 1n \log_2 \|f_{\e}\|_2 ~\le~ \max_{1 \le i \le r} \Big\{\phi_{\e}\(\alpha_i\) + \nu_i\Big\} \quad \mbox{and} \quad \frac 1n \log_2 \|f\|_q ~=~ \max_{1 \le i \le r} \Big\{\frac{\alpha_i - 1}{q} + \nu_i \Big\}.
\]

\noi Hence (\ref{ineq:NHC}) reduces to claim (\ref{ineq:NHC-phi}) in the following proposition.

\pro
\label{pro:NHC-phi}

\noi Let $q > 1$ and let $0 \le \alpha_1,...,\alpha_r \le 1$, $-1 \le \nu_1,...,\nu_r \le 1$ with $\max_{1 \le i \le r} \Big\{\(\alpha_i - 1\) + \nu_i\Big\} = 0$. Let $N = \max_{1 \le i \le r} \Big\{\frac{\alpha_i - 1}{q} + \nu_i\Big\}$. Then for any $0 \le \e \le \frac12$ holds
\beqn
\label{ineq:NHC-phi}
\max_{1 \le i \le r} \Big\{\phi_{\e}\(\alpha_i\) + \nu_i\Big\} ~\le~ \max_{1 \le i \le r} \Big\{\frac{\alpha_i-1}{\kappa} + \nu_i\Big\},
\eeqn
where $\kappa = \kappa_{2,q}\(\frac{qN}{q-1}, \e\)$ is defined in Definition~\ref{dfn:psi-kappa}.\footnote{It is easy to see that $0 \le N \le \frac{q-1}{q}$, and hence $\kappa$ is well defined.}

\noi Moreover, this is tight, in the following sense. For any $0 < N < \frac{q-1}{q}$ and $0 < \e < \frac12$, and for any $\tilde{\kappa} < \kappa_{2,q}(x,\e)$, there exist $0 \le \alpha_1,\alpha_2 \le 1 $ and $-1 \le \nu_1, \nu_2 \le 1$ such that $\max_{1 \le i \le 2} \Big\{\(\alpha_i - 1\) + \nu_i\Big\} = 0$, $\max_{1 \le i \le 2} \Big\{\frac{\alpha_i - 1}{q} + \nu_i\Big\} = N$, and $\max_{1 \le i \le 2} \Big\{\phi_{\e}\(\alpha_i\) + \nu_i\Big\} > \max_{1 \le i \le r} \Big\{\frac{\alpha_i-1}{\tilde{\kappa}} + \nu_i\Big\}$.
\epro

\prf (of Proposition~\ref{pro:NHC-phi}).

\noi  We start with verifying simple boundary cases. First, we observe that $\phi_0(x) = \frac{x-1}{2}$ (Lemma~\ref{lem:phi-derivatives}) and that $\phi_{\frac12}(x) = x-1$ (see the relevant discussion in the proof of Corollary~\ref{cor:useful}). In addition, it is easy to see that $\kappa_{2,q}\(x,\frac12\) = 1$ for all $q \ge 1$ and $0 \le x \le 1$; and (bearing in mind that $\phi_0(x) = \frac{x-1}{2}$) that $\kappa_{2,q}(x,0) = 2$ for all $q \ge 1$ and $0 \le x \le 1$. Therefore (\ref{ineq:NHC-phi}) is an identity for $\e = 0$ and $\e = \frac 12$. Hence we may and will assume from now on that $0 < \e < \frac12$.

\noi Let $q_0 = 1 + (1 - 2\e)^2$. We proceed to consider the (simple) cases $N = 0$ or $N + \frac 1q \le \frac{1}{q_0}$. Note that in these cases we have $\kappa = \kappa_{2,q}\(\frac{qN}{q-1}, \e\) = q_0$. Next, observe that, by the first and the second claims of Lemma~\ref{lem:phi-technical}, for any $0 \le \alpha \le 1$ holds $\phi_{\e}(\alpha) \le \frac{\alpha-1}{q_0} = \frac{\alpha-1}{\kappa}$ and hence (\ref{ineq:NHC-phi}) holds trivially in these cases.

\noi We continue to prove (\ref{ineq:NHC-phi}), assuming from now on that $N > 0$ and that $N + \frac 1q > \frac{1}{q_0}$. Let $\Omega \subseteq \R^2$ be the set defined in Definition~\ref{dfn:Omega}. We now define a family of continuous functions on $\Omega$, which will play an important role in the following argument. Let $\(\alpha_1, \nu_1\)$ be a point in $\Omega$ with $\frac{\alpha_1-1}{q} + \nu_1 = N$. Define a function $f = f_{\alpha_1, \nu_1}$ on $\Omega$ as follows. For $(\alpha, \nu) \in \Omega$ with $\alpha < 1$ let $f(\alpha, \nu)$ be the value of $\kappa$ for which $\phi_{\e}\(\alpha\) + \nu = \max\left\{\frac{\alpha_1-1}{\kappa} + \nu_1, \frac{\alpha-1}{\kappa} + \nu\right\}$. In addition, let $f(1, 0) = \frac{1-\alpha_1}{\nu_1}$.

\lem
\label{lem:f-cont}
For any choice of $\(\alpha_1, \nu_1\)$ as above the function $f_{\alpha_1,\nu_1}$ is well-defined and continuous on $\Omega$.
\elem

\noi Let $M\(\alpha_1, \nu_1\) = \max_{\Omega} f_{\alpha_1, \nu_1}$. The inequality (\ref{ineq:NHC-phi}) will follow from the next main technical claim, describing the behavior of $M\(\alpha_1, \nu_1\)$, as a function of $\alpha_1$ and $\nu_1$. Before stating this claim, let us make some preliminary comments. Note that the points $\(1 - \frac{qN}{q-1},\frac{qN}{q-1}\)$ and $\(0,N + \frac 1q\)$ are possible choices for $\(\alpha_1, \nu_1\)$. Note also that $\alpha_0$ in the third part of the claim is well-defined, by the fourth claim of Lemma~\ref{lem:phi-technical}.

\pro
\label{pro:main-tech}

\begin{enumerate}

\item
\[
M\(1 - \frac{qN}{q-1},\frac{qN}{q-1}\) ~=~ \frac{-\frac{qN}{q-1}}{\phi_{\e}\(1 - \frac{qN}{q-1}\)}.
\]

\item If $\frac{-\frac{qN}{q-1}}{\phi_{\e}\(1 - \frac{qN}{q-1}\)} \ge q$, then for any choice of $\(\alpha_1, \nu_1\)$ holds
\[
M\(\alpha_1, \nu_1\) ~\le~ M\(1 - \frac{qN}{q-1},\frac{qN}{q-1}\).
\]

\item If $\frac{-\frac{qN}{q-1}}{\phi_{\e}\(1 - \frac{qN}{q-1}\)} \le q$, then for any choice of $\(\alpha_1, \nu_1\)$ holds
\[
M\(1 - \frac{qN}{q-1},\frac{qN}{q-1}\) ~\le~ M\(\alpha_1, \nu_1\) ~\le~ M\(0,N + \frac 1q\) ~=~ \frac{\alpha_0 - 1}{\phi_{\e}\(\alpha_0\)},
\]
\[
\mathrm{where}~~\alpha_0~~ \mathrm{is~determined~by}~~ 1 - \alpha_0 - \frac{\alpha_0 \phi_{\e}\(\alpha_0\)}{1-\alpha_0} ~=~ N + \frac 1q.
\]

\end{enumerate}

\epro

\ignore{
\rem
\label{rem:meet}

\noi It follows that if $\frac{-\frac{qN}{q-1}}{\phi_{\e}\(1 - \frac{qN}{q-1}\)} = q$ then for any choice of $\(\alpha_1, \nu_1\)$ holds $M\(\alpha_1, \nu_1\) = q$.
\erem
}

\noi We will prove Lemma~\ref{lem:f-cont} and Proposition~\ref{pro:main-tech} below. For now we assume their validity and complete the proof of Proposition~\ref{pro:NHC-phi}.

%\prf (of Proposition~\ref{pro:NHC-phi}).

\noi We first prove (\ref{ineq:NHC-phi}). Note that if $x = \frac{qN}{q-1}$ then in the definition of $\kappa_{2,q}(x,\e)$ we have $y = \frac{q-1}{q} \cdot x + \frac 1q = N + \frac 1q$. Recall also that we may assume that $N > 0$ and that $y = N + \frac 1q > \frac{1}{q_0}$.

\noi By assumption $\alpha_i + \nu_i \le 1$, and $\frac{\alpha_i-1}{q} + \nu_i \le N$ for all $1 \le i \le r$ . Moreover there is an index $1 \le i \le r$ for which $\frac{\alpha_i-1}{q} + \nu_i = N$. Assume, w.l.o.g., that $i=1$.
We apply Proposition~\ref{pro:main-tech} to the function $f_{\alpha_1, \nu_1}$. Observe that the claim of the proposition together with the definition of $\kappa$ imply $M\(\alpha_1, \nu_1\) \le \kappa$. By the definition of $f_{\alpha_1, \nu_1}$, this means that for any point $\(\alpha, \nu\) \in \Omega$ holds $\phi_{\e}\(\alpha\) + \nu \le \max\left\{\frac{\alpha_1-1}{\kappa} + \nu_1, \frac{\alpha-1}{\kappa} + \nu\right\}$. We now claim that this inequality holds for all the points $\(\alpha_i, \nu_i\)$, $1 \le i \le r$, which will immediately imply (\ref{ineq:NHC-phi}). In fact, points $\(\alpha_i, \nu_i\)$ with $0 \le \nu_i \le 1$ lie in $\Omega$ and hence the inequality holds for these points. Furthermore, if $\nu_i < 0$ for some $1 \le i \le r$, then the point $\(\alpha_i, 0\)$ lies in $\Omega$, and hence $\phi_{\e}\(\alpha_i\) \le \max\left\{\frac{\alpha_1-1}{q} + \nu_1, \frac{\alpha_i-1}{q}\right\}$. But then $\phi_{\e}\(\alpha_i\) + \nu_i \le \max\left\{\frac{\alpha_1-1}{q} + \nu_1, \frac{\alpha_i-1}{q} + \nu_i\right\}$, proving the inequality in this case as well.

\noi We pass to proving the tightness of (\ref{ineq:NHC-phi}), starting with the case $N + \frac 1q \le \frac{1}{q_0}$. In this case, by definition,  $\kappa = q_0$. Let $\tilde{\kappa} < \kappa$ be given. Observe that since, by assumption, $N > 0$, we have $q > q_0$. Set $\alpha_1 = \frac{\frac{1}{q_0} - \frac 1q - N}{\frac{1}{q_0} - \frac 1q}$. Set $\nu_1 = \frac{1 - \alpha_1}{q_0}$. Let $\delta > 0$ be sufficiently small (depending on $N$ and $\tilde{\kappa}$). Set $\alpha_2 = 1 - \delta$ and $\nu_2 = \delta$. It is easy to see that $\alpha_1, \alpha_2$ and $\nu_1$, $\nu_2$ satisfy the required constraints. We claim that $\phi_{\e}\(\alpha_2\) + \nu_2 > \max_{1 \le i \le 2} \Big\{\frac{\alpha_i-1}{\tilde{\kappa}} + \nu_i\Big\}$. In fact, for a sufficiently small $\delta$ we have, using the second claim of Lemma~\ref{lem:phi-technical} (and observing that $\phi_{\e}'$ is continuous), that
\[
\phi_{\e}\(\alpha_2\) + \nu_2 ~=~ \phi_{\e}(1 - \delta) + \delta ~\approx~ -\frac{\delta}{q_0} + \delta ~>~ -\frac{\delta}{\tilde{\kappa}} + \delta ~=~ \frac{\alpha_2-1}{\tilde{\kappa}} + \nu_2,
\]
and
\[
\phi_{\e}\(\alpha_2\) + \nu_2 ~\approx~ -\frac{\delta}{q_0} + \delta ~>~ 0 ~\ge~ \frac{\alpha_1 - 1}{\tilde{\kappa}} + \frac{1 - \alpha_1}{q_0} ~=~ \frac{\alpha_1-1}{\tilde{\kappa}} + \nu_1.
\]

\noi We pass to the case $N + \frac 1q > \frac{1}{q_0}$ and $\frac{-\frac{qN}{q-1}}{\phi_{\e}\(1 - \frac{qN}{q-1}\)} \ge q$. In this case $\kappa = \frac{-\frac{qN}{q-1}}{\phi_{\e}\(1 - \frac{qN}{q-1}\)}$.  Set $\alpha_1 = \alpha_2 = 1 - \frac{qN}{q-1}$ and $\nu_1 = \nu_2 = \frac{qN}{q-1}$. It is easy to see that $\alpha_1, \alpha_2$ and $\nu_1$, $\nu_2$ satisfy the required constraints. It is also easy to see that for any $\tilde{\kappa} < \kappa$ holds
\[
\phi_{\e}\(\alpha_1\) + \nu_1 ~=~ \frac{\alpha_1 - 1}{\kappa} + \nu_1 ~>~ \frac{\alpha_1 - 1}{\tilde{\kappa}} + \nu_1.
\]

\noi It remains to deal with the case $N + \frac 1q > \frac{1}{q_0}$ and $\frac{-\frac{qN}{q-1}}{\phi_{\e}\(1 - \frac{qN}{q-1}\)} < q$. Let $\alpha_0$ be determined by $1 - \alpha_0 - \frac{\alpha_0 \phi_{\e}\(\alpha_0\)}{1-\alpha_0} = N + \frac 1q$. Then $\kappa = \frac{\alpha_0 - 1}{\phi_{\e}\(\alpha_0\)}$. Set $\alpha_1 = 0$ and $\nu_1 = N + \frac 1q$. Set $\alpha_2 = \alpha_0$ and $\nu_2 = 1 - \alpha_0$. It is easy to see that in this case the function $1 - \alpha - \frac{\alpha \phi_{\e}(\alpha)}{1 - \alpha}$ is larger than $N + \frac 1q$ at $\alpha = 1 - \frac{qN}{q-1}$, and hence the fourth claim of Lemma~\ref{lem:phi-technical} implies that $\alpha_2 = \alpha_0 > 1 - \frac{qN}{q-1}$. Using this, it is easy to see that $\alpha_1, \alpha_2$ and $\nu_1$, $\nu_2$ satisfy the required constraints. Furthermore, note that $\alpha_2 < 1$ (again, using the fourth claim of Lemma~\ref{lem:phi-technical}). It is also easy to verify, using the definition of $\alpha_0$, that
\[
\phi_{\e}\(\alpha_2\) + \nu_2 ~=~ \frac{\alpha_1 - 1}{\kappa} + \nu_1 ~=~ \frac{\alpha_2 - 1}{\kappa} + \nu_2,
\]
which implies that for any $\tilde{\kappa} < \kappa$ holds $\phi_{\e}\(\alpha_2\) + \nu_2 > \max_{1 \le i \le 2} \Big\{\frac{\alpha_i-1}{\tilde{\kappa}} + \nu_i\Big\}$. This completes the proof of Proposition~\ref{pro:NHC-phi}.

\eprf

\noi We now prove Lemma~\ref{lem:f-cont} and Proposition~\ref{pro:main-tech}. Recall that we may assume $N > 0$ and $N + \frac 1q > \frac{1}{q_0}$.

\prf (of Lemma~\ref{lem:f-cont}).

\noi Let $\(\alpha_1, \nu_1\)$ be a point in $\Omega$ with $\frac{\alpha_1-1}{q} + \nu_1 = N$. We start with some simple but useful observations about $\alpha_1$ and $\nu_1$.
\lem
\label{lem:alpha_1 and nu 1}
\begin{enumerate}

\item $\alpha_1 \le  1 - \frac{qN}{q-1}$ and $\nu_1 \ge \frac{qN}{q-1}$.

\item  $\frac{1 - \alpha_1}{\nu_1} < q_0$.

\end{enumerate}
\elem

\prf

\noi The first claim of the lemma is an easy consequence of the inequalities $\frac{\alpha_1-1}{q} + \nu_1 =  N$ and $\alpha_1 + \nu_1 \le 1$. We omit the details.

\noi We pass to the second claim of the lemma, distinguishing two cases, $q \le q_0$ and $q > q_0$. If $q \le q_0$, then $\nu_1 = N + \frac{1 - \alpha_1}{q} > \frac{1 - \alpha_1}{q} \ge \frac{1 - \alpha_1}{q_0}$. If $q > q_0$, we use the fact that $N + \frac 1q > \frac{1}{q_0}$ to obtain $\frac{1 - \alpha_1}{q_0} < \(1 - \alpha_1\) \(N + \frac 1q\) = \(1 - \alpha_1\) \(\frac{\alpha_1}{q} + \nu_1\)$. Viewing the last expression as a function of $\alpha_1$, it is easy to see that it equals $\nu_1$ at $\alpha_1 = 0$ and that it decreases in $\alpha_1$. Hence
$\nu_1 \ge \(1 - \alpha_1\) \(\frac{\alpha_1}{q} + \nu_1\) > \frac{1 - \alpha_1}{q_0}$, completing the argument in this case as well.
\eprf

\noi We now show that the function $f = f_{\alpha_1, \nu_1}$ is well-defined and that its values lie in the interval $\(0, q_0\)$. By Lemma~\ref{lem:alpha_1 and nu 1}, $\alpha_1 < 1$ and $ 0< f(1, 0) = \frac{1-\alpha_1}{\nu_1} < q_0$. Let now $\alpha < 1$. In this case the function $g(\kappa) = \max\left\{\frac{\alpha_1-1}{\kappa} + \nu_1, \frac{\alpha-1}{\kappa} + \nu\right\}$ is a strictly increasing continuous function of $\kappa$, which is $-\infty$ at $\kappa=0$. Furthermore, by Lemma~\ref{lem:phi-technical}, $\phi_{\e}(\alpha) < \frac{\alpha-1}{q_0}$, implying that $g\(q_0\) > \phi_{\e}\(\alpha\) + \nu$. Hence, by the intermediate value theorem, there exists a unique $0 < \kappa < q_0$ for which $\phi_{\e}(\alpha) + \nu = \max\left\{\frac{\alpha_1-1}{\kappa} + \nu_1, \frac{\alpha-1}{\kappa} + \nu\right\}$.

\noi Next, we argue that $f$ is continuous on $\Omega$. Let $\(\alpha, \nu\) \in \Omega$. If $\alpha < 1$, then there exists a compact neighborhood of $\(\alpha, \nu\)$ in which both one-sided derivatives of $g(\kappa)$ are positive and bounded. This, together with the fact that $\phi_{\e}(\alpha) + \nu$ is continuous, implies that $f$ is continuous at $\(\alpha, \nu\)$.

\noi It remains to argue that $f$ is continuous at $(1,0)$. Let $O$ be a sufficiently small neighbourhood of $(1, 0)$ in $\Omega$. Let $\(\alpha, \nu\) \in O$, with $\alpha < 1$. Then $\phi_{\e}(\alpha) + \nu$ is close to $\phi_{\e}(1) + 0 = 0$.  We would like to claim that $f(\alpha, \nu)$ is close to $f(1,0) = \frac{1 - \alpha_1}{\nu_1}$. In fact, assume towards contradiction that $f(\alpha, \nu)$ is significantly larger than $\frac{1 - \alpha_1}{\nu_1}$. In this case $\phi_{\e}(\alpha) + \nu = \max\left\{\frac{\alpha_1-1}{f(\alpha,\nu)} + \nu_1, \frac{\alpha-1}{f(\alpha,\nu)} + \nu\right\} \ge \frac{\alpha_1 - 1}{f(\alpha,\nu)} + \nu_1$ is significantly larger than $0$ (taking into account that $\alpha_1 < 1$), reaching contradiction. On the other hand, assume that $f(\alpha, \nu)$ is significantly smaller than $\frac{1 - \alpha_1}{\nu_1}$, and hence significantly smaller than $q_0$ (by the second claim of Lemma~\ref{lem:alpha_1 and nu 1}). Recall that $\phi_{\e}(1) = 0$ and that $\phi_{\e}'(1) = \frac{1}{q_0}$. Hence $\phi_{\e}\(\alpha\) = \frac{\alpha - 1}{q_0} + O\((1 - \alpha)^2\) > \frac{\alpha - 1}{f(\alpha,\nu)}$. This means that $\phi_{\e}(\alpha) + \nu = \frac{\alpha_1-1}{f(\alpha,\nu)} + \nu_1$, which is significantly smaller than $0$, again reaching contradiction. This completes the proof of Lemma~\ref{lem:f-cont}.

\eprf

\noi We collect some useful properties of $f = f_{\alpha_1, \nu_1}$ in the following claim.

\cor
\label{cor:separate}
\begin{enumerate}

\item For any $(\alpha, \nu) \in \Omega$ holds $\phi_{\e}(\alpha) + \nu = \max\left\{\frac{\alpha_1-1}{f(\alpha,\nu)} + \nu_1, \frac{\alpha-1}{f(\alpha,\nu)} + \nu\right\}$.

\item $0 < f \le M\(\alpha_1, \nu_1\) < q_0$ on $\Omega$.

\item For any $(\alpha, \nu) \in \Omega$ holds $f(\alpha, \nu) \le \frac{\alpha - 1}{\phi_{\e}(\alpha)}$. (If $\alpha=1$ we replace the RHS of this inequality with $q_0$.)

\end{enumerate}
\ecor

\prf
%The first two claims follow immediately from the preceding discussion and from the continuity of $f$. For the third claim, recall that $\phi_{\e}(\alpha) + \nu = \max\left\{\frac{\alpha_1-1}{f(\alpha,\nu)} + \nu_1, \frac{\alpha-1}{f(\alpha,\nu)} + \nu\right\} \ge \frac{\alpha-1}{f(\alpha,\nu)} + \nu$.
\eprf

\prf (Of Proposition~\ref{pro:main-tech})

\noi Let $\(\alpha_1, \nu_1\)$ be given, let $f = f_{\alpha_1, \nu_1}$, and let $M = M\(\alpha_1, \nu_1\) = \max_{\Omega} f$. Let $\(\alpha^{\ast}, \nu^{\ast}\)$ be a maximum point of $f$. Then $f\(\alpha^{\ast}, \nu^{\ast}\) = M$ and hence $\phi_{\e}\(\alpha^{\ast}\) + \nu^{\ast} = \max\left\{\frac{\alpha_1-1}{M} + \nu_1, \frac{\alpha^{\ast}-1}{M} + \nu\right\}$. Clearly either $\frac{\alpha_1 - 1}{M} + \nu_1 \not = \frac{\alpha^{\ast}-1}{M} + \nu^{\ast}$ or $\frac{\alpha_1 - 1}{M} + \nu_1 = \frac{\alpha^{\ast}-1}{M} + \nu^{\ast}$. In the first case we say that $\(\alpha^{\ast}, \nu^{\ast}\)$ is a maximum point of the {\it first type}, and otherwise it is a maximum point of the {\it second type}.

\noi The following two claims constitute the main steps of the proof of Proposition~\ref{pro:main-tech}. They describe the respective behavior of maxima points of the first and the second type.

\lem
\label{lem:first type}
\noi Let $\,\(\alpha^{\ast}, \nu^{\ast}\)$ be a maximum point of $\,f$ of the first type. Then the following two claims hold.

\begin{itemize}

\item  $\frac{\alpha_1 - 1}{f\(\alpha^{\ast}, \nu^{\ast}\)} + \nu_1 > \frac{\alpha^{\ast}-1}{f\(\alpha^{\ast}, \nu^{\ast}\)} + \nu^{\ast}$.

\item $\alpha^{\ast} \le 1 - \frac{qN}{q-1}$.

\end{itemize}
\elem

\lem
\label{lem:second type}

\noi If $\(\alpha_1, \nu_1\) = \(1 - \frac{qN}{q-1}, \frac{qN}{q-1}\)$, then $ \(1 - \frac{qN}{q-1}, \frac{qN}{q-1}\)$ is the unique maximum point of $f$. This is a maximum point of the second type.

\noi If $\(\alpha_1, \nu_1\) \not = \(1 - \frac{qN}{q-1}, \frac{qN}{q-1}\)$, then there are two possible cases.

\begin{itemize}

\item $\frac{-\frac{qN}{q-1}}{\phi_{\e}\(1 - \frac{qN}{q-1}\)} \ge q$. Let $\,\(\alpha^{\ast}, \nu^{\ast}\)$ be a maximum point of $\,f$ of the second type in this case. Then $\alpha^{\ast} \le 1 - \frac{qN}{q-1}$.

\item $\frac{-\frac{qN}{q-1}}{\phi_{\e}\(1 - \frac{qN}{q-1}\)} < q$. In this case $f$ has a unique maximum point $\,\(\alpha^{\ast}, \nu^{\ast}\)$. This point is of the second type. Furthermore, $\alpha^{\ast} > 1 - \frac{qN}{q-1}$, and it is uniquely determined by the following identity:
\[
\frac{\alpha^{\ast}-1}{\phi_{\e}(\alpha^{\ast})} ~=~ \frac{\alpha^{\ast} - \alpha_1}{\alpha^{\ast} - \(1 - \nu_1\)}.
\]

\end{itemize}
\elem

\noi Lemmas~\ref{lem:first type}~and~\ref{lem:second type} will be proved below. At this point we prove Proposition~\ref{pro:main-tech} assuming these lemmas hold.

\noi We start with the first claim of Proposition~\ref{pro:main-tech}. Let $\alpha_1 = 1 - \frac{qN}{q-1}$ and $\nu_1 = \frac{qN}{q-1}$. Let $f = f_{\alpha_1, \nu_1}$. By the first claim of Lemma~\ref{lem:second type}, we have
\[
M\(\alpha_1, \nu_1\) ~=~ f\(\alpha_1, \nu_1\) ~=~ \frac{\alpha_1 - 1}{\phi_{\e}\(\alpha_1, \nu_1\)} ~=~ \frac{-\frac{qN}{q-1}}{\phi_{\e}\(1 - \frac{qN}{q-1}\)}.
\]

\noi We pass to the second claim of the proposition. Assume that $\frac{-\frac{qN}{q-1}}{\phi_{\e}\(1 - \frac{qN}{q-1}\)} \ge q$. Let $f = f_{\alpha_1, \nu_1}$, for some $\alpha_1$ and $\nu_1$. Let $\(\alpha^{\ast}, \nu^{\ast}\)$ be a maximum point of $f$. Then  Lemmas~\ref{lem:first type}~and~\ref{lem:second type} imply that $\alpha^{\ast} \le 1 - \frac{qN}{q-1}$. Hence
\[
M\(\alpha_1, \nu_1\) ~=~ f\(\alpha^{\ast}, \nu^{\ast}\) ~\le~ \frac{\alpha^{\ast} - 1}{\phi_{\e}\(\alpha^{\ast}\)} ~\le~ \frac{-\frac{qN}{q-1}}{\phi_{\e}\(1 - \frac{qN}{q-1}\)} ~=~ M\(1 - \frac{qN}{q-1},\frac{qN}{q-1}\).
\]
Here in the second step we have used the third claim of Corollary~\ref{cor:separate}, in the third step the third claim of Lemma~\ref{lem:phi-technical} and in the fourth step the first claim of the proposition.

\noi We pass to the third claim of the proposition.  Assume that $\frac{-\frac{qN}{q-1}}{\phi_{\e}\(1 - \frac{qN}{q-1}\)} < q$. Let $f = f_{\alpha_1, \nu_1}$, for some $\alpha_1$ and $\nu_1$. Then, by Lemma~\ref{lem:second type}, $f$ has a unique maximum point $\(\alpha^{\ast}, \nu^{\ast}\)$. This means that $\alpha^{\ast}$ is determined by $\alpha_1$ and $\nu_1$, and furthermore, since $\nu_1 = N + \frac{1 - \alpha_1}{q}$, $\alpha^{\ast}$ is a function of $\alpha_1$. We will show the following claim below.

\lem
\label{lem:alpha-ast}
If $\(\alpha_1, \nu_1\) \not = \(1 - \frac{qN}{q-1}, \frac{qN}{q-1}\)$ and $\frac{-\frac{qN}{q-1}}{\phi_{\e}\(1 - \frac{qN}{q-1}\)} < q$, then $\alpha^{\ast}$ is a decreasing function of $\alpha_1$.
\elem

\noi Assume Lemma~\ref{lem:alpha-ast} to hold. We have
\[
M\(\alpha_1, \nu_1\) ~=~ f\(\alpha^{\ast}, \nu^{\ast}\) ~=~ \frac{\alpha^{\ast}\(\alpha_1\) - 1}{\phi_{\e}\(\alpha^{\ast}\(\alpha_1\)\)} ~\le~ \frac{\alpha^{\ast}\(0\) - 1}{\phi_{\e}\(\alpha^{\ast}\(0\)\)} ~=~ M\(0,N + \frac 1q\).
\]
The second step uses the fact that $\(\alpha^{\ast}, \nu^{\ast}\)$ is a maximum point of the second type, and hence $f\(\alpha^{\ast}, \nu^{\ast}\) = \frac{\alpha^{\ast} - 1}{\phi\(\alpha^{\ast}\)}$. The third step uses Lemma~\ref{lem:alpha-ast} and the third claim of Lemma~\ref{lem:phi-technical}, and the fourth step the fact that $\alpha_1 = 0$ implies $\nu_1 = N + \frac 1q$.

\noi Next, by Lemma~\ref{lem:second type}, $\alpha = \alpha^{\ast}\(0\)$ is determined by the identity $\frac{\alpha -1}{\phi_{\e}(\alpha)} ~=~ \frac{\alpha}{\alpha  - \(\frac{q-1}{q} - N\)}$ which, after rearranging, gives $1 - \alpha - \frac{\alpha \phi_{\e}(\alpha)}{1-\alpha} ~=~ N + \frac 1q$. Hence, by the fourth claim of Lemma~\ref{lem:phi-technical}, $\alpha^{\ast}\(0\) = \alpha_0$ and $M\(0,N + \frac 1q\) = \frac{\alpha_0 - 1}{\phi_{\e}\(\alpha_0\)}$.

\noi To conclude the proof of the third claim of the proposition, observe that since $\alpha^{\ast} > 1 - \frac{qN}{q-1}$, we have
\[
M\(\alpha_1, \nu_1\) ~=~ f\(\alpha^{\ast}, \nu^{\ast}\) ~=~ \frac{\alpha^{\ast} - 1}{\phi_{\e}\(\alpha^{\ast}\)} ~>~ \frac{-\frac{qN}{q-1}}{\phi_{\e}\(1 - \frac{qN}{q-1}\)},
\]
where the last inequality is by the third claim of Lemma~\ref{lem:phi-technical}. This completes the proof of Proposition~\ref{pro:main-tech}.
\eprf

\noi It remains to prove Lemmas~\ref{lem:first type},~\ref{lem:second type},~and~\ref{lem:alpha-ast}.

\prf (Of Lemma~\ref{lem:first type})

\noi We start with the first claim of the lemma. Assume towards contradiction that $\frac{\alpha_1 - 1}{f\(\alpha^{\ast}, \nu^{\ast}\)} + \nu_1 < \frac{\alpha^{\ast}-1}{f\(\alpha^{\ast}, \nu^{\ast}\)} + \nu^{\ast}$. Since $f$ is a positive continuous function on $\Omega$, there is a neighborhood $O$ of $\(\alpha^{\ast}, \nu^{\ast}\)$ in $\Omega$ on which $\frac{\alpha_1 - 1}{f\(\alpha, \nu\)} + \nu_1 < \frac{\alpha - 1}{f\(\alpha, \nu\)} + \nu$. This means that any point $(\alpha, \nu) \in O$ satisfies $\phi_{\e}\(\alpha\) + \nu = \frac{\alpha - 1}{f\(\alpha, \nu\)} + \nu$, and hence $f(\alpha, \nu) = \frac{\alpha-1}{\phi_{\e}(\alpha)}$. Since $f\(\alpha^{\ast}, \nu^{\ast}\) \ge f\(\alpha, \nu\)$, this implies that $\frac{\alpha^{\ast}-1}{\phi_{\e}(\alpha^{\ast})} \ge \frac{\alpha-1}{\phi_{\e}(\alpha}$, and hence, by the third claim of Lemma~\ref{lem:phi-technical}, that $\alpha^{\ast} \ge \alpha$. It follows that $\alpha^{\ast}$ has to be $1$, and hence $\(\alpha^{\ast}, \nu^{\ast}\) = (1,0)$. But in this case $\frac{\alpha_1 - 1}{f\(\alpha^{\ast}, \nu^{\ast}\)} + \nu_1 = \frac{\alpha^{\ast}-1}{f\(\alpha^{\ast}, \nu^{\ast}\)} + \nu^{\ast} = 0$, reaching contradiction.

\noi We pass to the second claim of the lemma. By the first claim $\frac{\alpha_1 - 1}{f\(\alpha^{\ast}, \nu^{\ast}\)} + \nu_1 > \frac{\alpha^{\ast}-1}{f\(\alpha^{\ast}, \nu^{\ast}\)} + \nu^{\ast}$. We claim that this implies that $\(\alpha^{\ast}, \nu^{\ast}\)$ is a local maximum of $\phi_{\e}(\alpha) + \nu$. In fact, arguing as above, there is a neighborhood $O$ of $\(\alpha^{\ast}, \nu^{\ast}\)$ on which $\frac{\alpha_1 - 1}{f\(\alpha, \nu\)} + \nu_1 > \frac{\alpha - 1}{f\(\alpha, \nu\)} + \nu$. This means that for any point $(\alpha, \nu) \in O$ we have $\phi_{\e}\(\alpha\) + \nu = \frac{\alpha_1 - 1}{f\(\alpha, \nu\)} + \nu_1$. Since $f\(\alpha^{\ast}, \nu^{\ast}\) \ge f\(\alpha, \nu\)$, this implies that $\phi_{\e}(\alpha) + \nu \le \phi\(\alpha^{\ast}\) + \nu^{\ast}$. To complete the proof, recall that any local maximum $(\alpha, \nu)$ of $\phi(\alpha) + \nu$ has $\alpha \le 1 - \frac{qN}{q-1}$ (as shown in the proof of Proposition~\ref{pro:MrsG-2p}).
\eprf

\prf (Of Lemma~\ref{lem:second type})

\noi Let $\(\alpha^{\ast}, \nu^{\ast}\)$ be a maximum point of $f$ of the second type. The first observation is that $\(\alpha^{\ast}, \nu^{\ast}\)$ has to lie on the upper boundary of $\Omega$. In fact, assume not. Then for a sufficiently small $\tau > 0$ the point $\(\alpha, \nu\) = \(\alpha^{\ast}, \nu^{\ast} + \tau\)$ is in $\Omega$. Since $f\(\alpha, \nu\) \le f\(\alpha^{\ast}, \nu^{\ast}\)$, we have $\phi_{\e}\(\alpha\) + \nu > \phi_{\e}\(\alpha^{\ast}\) + \nu^{\ast} = \frac{\alpha_1 - 1}{f\(\alpha^{\ast}, \nu^{\ast}\)} + \nu_1 \ge \frac{\alpha_1 - 1}{f\(\alpha, \nu\)} + \nu_1$. Hence $f\(\alpha, \nu\)$ is determined by the equality $\phi_{\e}\(\alpha\) + \nu = \frac{\alpha-1}{f\(\alpha, \nu\)} + \nu$, which implies $f\(\alpha, \nu\) = f\(\alpha^{\ast}, \nu^{\ast}\) = \frac{\alpha^{\ast}-1}{\phi_{\e}\(\alpha^{\ast}\)}$. Hence $\(\alpha, \nu\)$ is a point of maximum of $f$ of the first type with $\frac{\alpha_1 - 1}{f\(\alpha, \nu\)} + \nu_1 < \frac{\alpha-1}{f\(\alpha, \nu\)} + \nu$. This, however, contradicts the first claim of Lemma~\ref{lem:first type}.

\noi Recall that the upper boundary of $\Omega$ is a piecewise linear curve which starts as the straight line $\frac{\alpha}{q} + \nu = N + \frac 1q$, for $0 \le \alpha \le 1 - \frac{qN}{q-1}$ and continues as the straight line $\alpha + \nu = 1$ for $1 - \frac{qN}{q-1} \le \alpha \le 1$. Hence there are two cases to consider: In the first case $\alpha^{\ast} \le 1 - \frac{qN}{q-1}$ and $\frac{\alpha^{\ast}}{q} + \nu^{\ast} = N + \frac 1q$. In the second case $1 - \frac{qN}{q-1} < \alpha^{\ast} \le 1$ and $\alpha^{\ast} + \nu^{\ast} = 1$.

\noi Assume that the second case holds. Then $\(\alpha^{\ast}, \nu^{\ast}\)$ satisfies

\begin{enumerate}

\item $\frac{\alpha_1 - 1}{f\(\alpha^{\ast},\nu^{\ast}\)} + \nu_1  = \frac{\alpha^{\ast}-1}{f\(\alpha^{\ast},\nu^{\ast}\)} + \nu^{\ast} = \phi_{\e}\(\alpha^{\ast}\) + \nu^{\ast}$.

\item $1 - \frac{qN}{q-1} < \alpha^{\ast}  \le 1 $ and $\alpha^{\ast}  + \nu^{\ast} = 1$.

\end{enumerate}

\noi In particular, $f\(\alpha^{\ast},\nu^{\ast}\) = \frac{\alpha^{\ast}-1}{\phi_{\e}\(\alpha^{\ast}\)} = \frac{\alpha^{\ast} - \alpha_1}{\alpha^{\ast} - \(1 - \nu_1\)}$. Consider the following two functions of $\alpha$: $g_1(\alpha) = \frac{\alpha-1}{\phi_{\e}(\alpha)}$ and $g_2\(\alpha\) = \frac{\alpha - \alpha_1}{\alpha - \(1 - \nu_1\)}$, for $\alpha > 1 - \frac{qN}{q-1}$. Note that $g_2$ is well-defined since, by Lemma~\ref{lem:alpha_1 and nu 1}, $\nu_1 \ge \frac{qN}{q-1}$. By the third claim of Lemma~\ref{lem:phi-technical}, $g_1$ is strictly increasing. On the other hand, $g_2(\alpha) = 1 + \frac{1 - \alpha_1 - \nu_1}{\alpha - \(1 - \nu_1\)}$ is non-increasing. Note also that $g_1(1) = q_0$ (more precisely, $\lim_{\alpha \rarrow 1} g_1(\alpha) = q_0$) and, by Lemma~\ref{lem:alpha_1 and nu 1}, $g_2(1) = \frac{1 - \alpha_1}{\nu_1} < q_0$. This means that $g_1$ and $g_2$ coincide at a (unique) point $1 - \frac{qN}{q-1} < \alpha < 1$ iff $g_1\(1 - \frac{qN}{q-1}\) < g_2\(1 - \frac{qN}{q-1}\)$.

\noi Observe that if $\(\alpha_1, \nu_1\) = \(1 - \frac{qN}{q-1}, \frac{qN}{q-1}\)$ then $g_2$ is the constant $1$-function. Furthermore, by the first and the third claims of Lemma~\ref{lem:phi-technical}, $g_1\(1 - \frac{qN}{q-1}\) \ge g_1(0) =
\frac{2}{\log_2\(4/q_0\)} \ge 1$, and hence in this case $g_1$ and $g_2$ cannot coincide for $\alpha > 1 - \frac{qN}{q-1}$. If $\(\alpha_1, \nu_1\) \not = \(1 - \frac{qN}{q-1}, \frac{qN}{q-1}\)$ then it is easy to see (recall that $\frac{\alpha_1}{q} + \nu_1 = N + \frac{1}{q}$) that $ g_2\(1 - \frac{qN}{q-1}\) = q$, and hence the two functions have a unique intersection at some $\alpha > 1 - \frac{qN}{q-1}$ iff $g_1\(1 - \frac{qN}{q-1}\) = \frac{-\frac{qN}{q-1}}{\phi_{\e}\(1 - \frac{qN}{q-1}\)}$ is smaller than $q$.

\noi To recap, the second case can hold only provided $\(\alpha_1, \nu_1\) \not = \(1 - \frac{qN}{q-1}, \frac{qN}{q-1}\)$ and $\frac{-\frac{qN}{q-1}}{\phi_{\e}\(1 - \frac{qN}{q-1}\)} < q$. Furthermore, if it holds then $1 - \frac{qN}{q-1} < \alpha^{\ast} < 1$ is uniquely determined by the equality $g_1\(\alpha^{\ast}\) = g_2\(\alpha^{\ast}\)$.

\noi We can now complete the proof of the lemma. First, let $\(\alpha_1, \nu_1\) = \(1 - \frac{qN}{q-1}, \frac{qN}{q-1}\)$. By the preceding discussion, in this case a maximum point $\(\alpha^{\ast},\nu^{\ast}\)$ of $f$ of the second type has to have $\alpha^{\ast} \le \alpha_1$. Moreover, taking into account Lemma~\ref{lem:first type}, this is true for any maximum point of $f$. By the third claim of Corollary~\ref{cor:separate}, this means that $M\(\alpha_1, \nu_1\) \le \frac{\alpha_1 - 1}{\phi_{\e}\(\alpha_1\)} = f\(\alpha_1, \nu_1\)$. Hence $\(\alpha_1, \nu_1\)$ is a maximum point of $f$. It is trivially a maximum point of the second type. To see that it is a unique maximum point, note that for any point $(\alpha, \nu)$ on the upper boundary of $\Omega$, if $\alpha = \alpha_1$, then necessarily $\nu = \nu_1$. So, for any other putative maximum point $(\alpha, \nu)$, we would have $\alpha < \alpha_1$ and hence, by the third claims of Lemma~\ref{lem:phi-technical} and the third claim of Corollary~\ref{cor:separate}, $f(\alpha, \nu) \le \frac{\alpha - 1}{\phi_{\e}\(\alpha\)} < \frac{\alpha_1 - 1}{\phi_{\e}\(\alpha_1\)} = f\(\alpha_1, \nu_1\)$. This proves the first claim of the lemma.

\noi Assume now that $\(\alpha_1, \nu_1\) \not = \(1 - \frac{qN}{q-1}, \frac{qN}{q-1}\)$. Let $\(\alpha^{\ast}, \nu^{\ast}\)$ be a maximum point of $f$ of the second type. If $g_1\(1 - \frac{qN}{q-1}\) = \frac{-\frac{qN}{q-1}}{\phi_{\e}\(1 - \frac{qN}{q-1}\)} \ge q$, then the preceding discussion implies that $\alpha^{\ast} \le 1 - \frac{qN}{q-1}$, proving the second claim of the lemma.

\noi If $\frac{-\frac{qN}{q-1}}{\phi_{\e}\(1 - \frac{qN}{q-1}\)} < q$, let $\alpha$ be the unique solution for $g_1(\alpha) = g_2(\alpha)$ on $1 - \frac{qN}{q-1} < \alpha < 1$. Set $\alpha^{\ast} = \alpha$ and $\nu^{\ast} = 1 - \alpha$. We claim that $\(\alpha^{\ast},\nu^{\ast}\)$ is the unique maximum point of $f$ (note that by Lemma~\ref{lem:first type} it would necessarily be of the second type). In fact, let us first verify that $\frac{\alpha_1 - 1}{\kappa} + \nu_1  = \frac{\alpha^{\ast}-1}{\kappa} + \nu^{\ast} = \phi_{\e}\(\alpha^{\ast}\) + \nu^{\ast}$, for $\kappa = \frac{\alpha^{\ast} - 1}{\phi_{\e}\(\alpha^{\ast}\)}$. The second equality is immediate, by the definition of $\kappa$. The first equality is equivalent to $\kappa = \frac{\alpha^{\ast} - \alpha_1}{\alpha^{\ast} - \(1 - \nu_1\)}$, which follows from the definitions of $\alpha^{\ast}$ and $\kappa$. Hence $f\(\alpha^{\ast},\nu^{\ast}\) = \kappa = \frac{\alpha^{\ast} - 1}{\phi_{\e}\(\alpha^{\ast}\)}$. For any other putative maximum point $(\alpha, \nu)$, we would have, by the preceding discussion, that  $\alpha \le 1 - \frac{qN}{q-1} < \alpha^{\ast}$ and hence, as above, $f(\alpha, \nu) \le \frac{\alpha - 1}{\phi_{\e}\(\alpha\)} < f\(\alpha^{\ast}, \nu^{\ast}\)$. This proves the third claim of the lemma.
\eprf

\prf (of Lemma~\ref{lem:alpha-ast}).

\noi In the assumptions of the lemma, $\alpha^{\ast}$ is the unique solution on $\(1 - \frac{qN}{q-1},1\)$ of the identity
\[
\frac{\alpha^{\ast}-1}{\phi_{\e}(\alpha^{\ast})} ~=~ \frac{\alpha^{\ast} - \alpha_1}{\alpha^{\ast} - \(1 - \nu_1\)}.
\]
Here the LHS is a strictly increasing and the RHS a strictly decreasing (since by assumption $\alpha_1 \not = 1 - \frac{qN}{q-1}$, and hence $\alpha_1 + \nu_1 < 1$) functions of $\alpha^{\ast}$. It follows that to prove the claim of the lemma it suffices to show that for a fixed $\alpha^{\ast} > 1 - \frac{qN}{q-1}$ the RHS is a decreasing function of $\alpha_1$ (keeping in mind that $\nu_1 = -\frac{\alpha_1}{q} + \(N + \frac 1q\)$). But this is easily verifiable by a direct differentiation of the RHS w.r.t. $\alpha_1$.
\eprf

\noi This completes the proof of Proposition~\ref{pro:NHC-phi} and of (\ref{ineq:NHC}). We proceed to complete the proof of Theorem~\ref{thm:NHC}. The tightness of (\ref{ineq:NHC}) follows from the tightness of (\ref{ineq:NHC-phi}), similarly to the way the tightness of (\ref{ineq:MrsG-2p}) was shown in the proof of Proposition~\ref{pro:MrsG-2p}. We omit the details.

\noi It remains to consider the properties of the function $\kappa_{2,q}$. We first remark that it is easy to see, using the properties of the function $\phi_{\e}$ given in Lemma~\ref{lem:phi-technical}, that $\kappa_{2,q}$ is a continuous function of its first variable (we omit the details). In particular, we can replace strict inequalities with non-strict ones in the definition of $\kappa_{2,q}$ in Definition~\ref{dfn:psi-kappa}. Now there are two cases to consider.

\begin{itemize}

\item $q \ge q_0$. In this case, by the third claim of Lemma~\ref{lem:phi-technical}, $-\frac{x}{\phi_{\e}(1 - x)}$ is never larger than $q$, and hence
\[
\kappa_{2,q}(x,\e) ~=~ \left\{\begin{array}{ccc} q_0 & \mbox{if} & y \le \frac{1}{q_0} \\
\frac{\alpha_0 - 1}{\phi_{\e}(\alpha_0)} & \mbox{if} & y  \ge  \frac{1}{q_0} \end{array}\right.
\]
Here $y = \frac{q-1}{q} \cdot x + \frac 1q$, $q_0 = 1 + (1-2\e)^2$, and $\alpha_0$ is determined by $1 - \alpha_0 - \frac{\alpha_0 \phi_{\e}(\alpha_0)}{1-\alpha_0} = y$. Note that $\alpha_0$ is well-defined, by the fourth claim of Lemma~\ref{lem:phi-technical}. The fact that $\kappa_{2,q}$ is decreasing in $x$ follows from combining the third and the fourth claims of Lemma~\ref{lem:phi-technical}. In fact, $\kappa_{2,q}$ is a constant-$\(1 + (1-2\e)^2\)$ function for $0 \le x \le \frac{q-q_0}{(q-1)q_0}$, and it is strictly decreasing for larger $x$.

\item $q < q_0$. In this case $y$ is always greater than $\frac{1}{q_0}$ and we have that
\[
\kappa_{2,q}(x,\e) ~=~ \left\{\begin{array}{ccc} -\frac{x}{\phi_{\e}(1 - x)} & \mbox{if} &  -\frac{x}{\phi_{\e}(1 - x)} \ge q \\
\frac{\alpha_0 - 1}{\phi_{\e}(\alpha_0)} & \mbox{if} &  -\frac{x}{\phi_{\e}(1 - x)} \le q \end{array}\right.
\]
It suffices to show that $\kappa_{2,q}$ is decreasing on both relevant subintervals of $[0,1]$, and this again follows from the third and the fourth claims of Lemma~\ref{lem:phi-technical}. In this case $\kappa_{2,q}$ is strictly decreasing on $[0,1]$.

\end{itemize}

\section{Remaining proofs}
\label{sec:remaining}

\subsubsection*{Proof of Lemma~\ref{lem:phi-technical}}

\prf
The strict concavity of $\phi_{\e}$ and the bounds on its derivative were shown in \cite{KS2}, Lemma~2.13 (note that $\phi_{\e}(x) = \frac12 \tilde{\phi}(x,2\e(1-\e))$ in terms of \cite{KS2}). The value of $\phi_{\e}$ at the endpoints of the interval $[0,1]$ are directly computable.

\noi We pass to the third claim of the lemma. Taking the derivative and rearranging, it suffices to prove that for any $\alpha \in (0,1)$ holds $\phi_{\e}(\alpha) > (\alpha - 1) \phi'_{\e}(\alpha)$. This follows immediately from the strict concavity of $\phi_{\e}$ and the fact that $\phi_{\e}(1) = 0$.

\noi We pass to the last claim of the lemma. Taking the derivative and rearranging, it suffices to prove that for any $\alpha \in (0,1)$ holds
\[
(1-\alpha)\(\alpha \phi_{\e}'(\alpha) + (1-\alpha)\) > -\phi_{\e}(\alpha).
\]

\noi Since $(1-\alpha) \cdot \phi_{\e}'(\alpha) > -\phi_{\e}(\alpha)$, it suffices to show that $\alpha \phi_{\e}'(\alpha) + (1-\alpha) \ge \phi_{\e}'(\alpha)$, and this follows from the first two claims of the lemma. The values of the function $g$ at the endpoints are directly computable.
\eprf

\subsubsection*{Proof of Corollary~\ref{cor:Lsob}}

\prf
Let $q = 2$ and $\kappa = \kappa_{2,2}$ (see the second claim of Corollary~\ref{cor:useful} for a more explicit statement of Theorem~\ref{thm:NHC} in this case). Viewing both sides of (\ref{ineq:NHC}) as functions of $\e$, and writing $L(\e)$ for the LHS and $R(\e)$ for the RHS, we have $L(0) = R(0) = \|f\|_2$, and $L(\e) \le R(\e)$ for $0 \le \e \le \frac12$. It is easy to see that both $L$ and $R$ are differentiable, and we may deduce that $F'(0) \le G'(0)$. Computing the derivatives (see e.g., \cite{Gross}) gives
\[
F'(0) ~=~ -\frac12 \cdot \frac{{\cal E}(f,f)}{\|f\|_2} \quad \mbox{and} \quad G'(0) ~=~ \frac{\ln(2) \kappa'(0)}{4} \cdot \frac{Ent\(f^2\)}{\|f\|_2},
\]
where we write $\kappa'(0)$ for $\frac{\partial \kappa}{\partial \e}_{|\e = 0}$. Hence $F'(0) \le G'(0)$ is equivalent to
\beqn
\label{ineq:LS-aux}
{\cal E}(f,f) ~\ge~ -\frac{\ln(2) \kappa'(0)}{2} \cdot Ent\(f^2\).
\eeqn
We proceed to compute $\kappa'(0)$, starting with a technical lemma which deals with the behavior of the function $\phi_{\e}(x)$ around $\e = 0$. We again use the fact that $\phi_{\e}(x) = \Phi(x,2\e(1-\e)) = \frac12 \tilde{\phi}(x,2\e(1-\e))$, where the function $\tilde{\phi}$ was defined and studied in \cite{KS2}. In the calculations below $\phi(x,\e)$ is written instead of $\phi_{\e}(x)$, for notational convenience.

\lem
\label{lem:phi-derivatives}
\begin{enumerate}
  \item
  \[
  \phi(x,0) ~=~ \frac{x-1}{2}.
  \]

  \item
  \[
  \frac{\partial \phi}{\partial \e}\(x,0\) ~=~ \frac{2 \sqrt{H^{-1}(x) \(1 - H^{-1}(x)\)} - 1}{\ln(2)}.
  \]
\end{enumerate}
\elem

\prf

\noi Recall that
\[
\tilde{\phi}(x,\e) ~=~ x - 1 + \sigma H\(\frac{y}{\sigma}\) + (1 - \sigma) H\(\frac{y}{1-\sigma}\) + 2y \log_2(\e) + (1-2y)\log_2(1 - \e),
\]
where $\sigma = H^{-1}(x)$ and $y = y(x, \e) = \frac{-\e^2 + \e \sqrt{\e^2 + 4(1-2\e) \sigma (1-\sigma)}}{2(1-2\e)}$.

\noi The first claim of the lemma is verified by inspection, observing that $y(x,0) = 0$ for any $0 \le x \le 1$.

\noi We pass to the second claim of the lemma. Using (as in the proof of Lemma~2.13 in \cite{KS2}), the fact that for $\e > 0$ holds $\frac{(\sigma-y)(1 - \sigma - y)}{y^2} = \frac{(1-\e)^2}{\e^2}$, and writing $\delta = 2\e(1-\e)$, we have that
\[
\frac{\partial \phi(x,\e)}{\partial \e} ~=~ \frac12 \cdot \frac{\partial \tilde{\phi}(x,\delta)}{\partial \e} ~=~ \frac{1-2\e}{\ln(2)} \cdot \frac{2y - \delta}{\delta(1-\delta)}.
\]

\noi Hence
\[
\frac{\partial \phi(x,\e)}{\partial \e}_{| \e = 0} ~=~ \lim_{\e \rarrow 0} \frac{\partial \phi(x,\e)}{\partial \e} ~=~ \lim_{\delta \rarrow 0} \frac{\partial \phi(x,\e)}{\partial \e} ~=~
\]
\[
\frac{1}{\ln(2)} \cdot \(\lim_{\e \rarrow 0} \frac{2y}{\delta} - 1\) ~=~ \frac{2\sqrt{\sigma(1-\sigma)} - 1}{\ln(2)} ~=~ \frac{2 \sqrt{H^{-1}(x) \(1 - H^{-1}(x)\)} - 1}{\ln(2)}.
\]
This proves the second claim of the lemma.
\eprf

\noi In the following we will assume (as we may, since for constant functions the claim of the corollary is trivially true) that $f$ is not a constant function, implying that $x = \frac 1n Ent_2\(\frac{f}{\|f\|_1}\) > 0$. For $\e$ sufficiently close to zero, we have that $\frac{x+1}{2} > \frac{1}{q_0}$ (recall that $q_0 = 1 + (1-2\e)^2$) and hence $\kappa = \frac{\alpha - 1}{\phi(\alpha,\e)}$, where $\alpha = \alpha(\e)$ is determined by $1 - \alpha + \frac{\alpha \phi(\alpha, \e)}{\alpha - 1} = \frac{x+1}{2}$. Taking the derivative w.r.t. $\e$ in the definition of $\alpha$ and rearranging gives
\[
\alpha'(\e) ~=~ -\frac{\alpha(\alpha-1) \frac{\partial \phi}{\partial \e}(\alpha,\e)}{\alpha(\alpha-1)\frac{\partial \phi}{\partial \alpha}(\alpha,\e) - \phi(\alpha,\e) - (\alpha-1)^2}.
\]

\noi Using the first claim of Lemma~\ref{lem:phi-derivatives}, it is easy to see that $\alpha(0) = 1 - x$. Hence, using both claims of Lemma~\ref{lem:phi-derivatives}, we have that
\[
\alpha'(0) ~=~ \lim_{\e \rarrow 0} \alpha'(\e) ~=~ -\frac{2}{\ln 2} \cdot \frac{\alpha(0) \(2 \sqrt{H^{-1}\(\alpha(0)\) \(1 - H^{-1}\(\alpha(0)\)\)} - 1\)}{1 - \alpha(0)} ~=~
\]
\[
-\frac{2}{\ln 2} \cdot \frac{(1-x) \(2 \sqrt{H^{-1}\(1 - x\) \(1 - H^{-1}\(1 - x\)\)} - 1\)}{x}
\]

\noi Next, observe that by the definition of $\kappa$, we have that $1 - \alpha + \frac{\alpha}{\kappa} = \frac{x+1}{2}$. In particular, $\kappa(0) = 2$. Taking the derivative of this identity w.r.t. $\e$ and rearranging, we get
\[
\kappa'(0) ~=~ -\frac{2\alpha'(0)}{\alpha(0)} ~=~ \frac{4}{\ln 2} \cdot \frac{\(2 \sqrt{H^{-1}\(1 - x\) \(1 - H^{-1}\(1 - x\)\)} - 1\)}{x} ~=~
\]
\[
-\frac{2}{\ln(2)} C(x) ~=~ -\frac{2}{\ln(2)} C\(\frac{Ent_2\(\frac{f}{\|f\|_1}\)}{n}\).
\]
Combining this with (\ref{ineq:LS-aux}) gives
\[
{\cal E}(f,f) ~\ge~ C\(\frac{Ent_2\(\frac{f}{\|f\|_1}\)}{n}\) \cdot Ent\(f^2\),
\]
as claimed. The fact that $C(\cdot)$ a convex and increasing function on $[0,1]$, taking $[0,1]$ onto $\left[2 \ln 2, 2\right]$ was proved in \cite{Modified Log-Sobolev}.

\eprf

\subsubsection*{Proof of Corollary~\ref{cor:useful}}

\prf

\noi We start with the first claim of the corollary. First consider the case $\e = \frac 12$. It is easy to see that $\phi_{\frac12}(x) = x-1$ (note that in the definition of $\Phi(x,\e)$ we have $y\(x, \frac12\) = \lim_{\e \rarrow \frac12} y\(x, \e\) = H^{-1}(x)\(1-H^{-1}(x)\)$) and hence in this case the value of $\kappa$ given by the claim is $1$ (as it should be).

\noi Assume now $\e < \frac12$. This implies that $q_0 = 1 + (1 - 2\e)^2 > 1$. By the first claim of Lemma~\ref{lem:phi-technical}, this means that for any $0 \le x \le 1$ we have $\frac{-x}{\phi_{\e}(1-x)} \ge -\frac{1}{\phi_{\e}(0)} = \frac{2}{\log_2\(\frac{4}{q_0}\)} > 1$. Hence, it is easy to see that for $q$ sufficiently close to $1$ the first and the third clauses in the definition of $\kappa_{2,q}$ in Definition~\ref{dfn:psi-kappa} do not apply, and we have $\kappa_{2,q}(x,\e) = \frac{-x}{\phi_{\e}(1-x)}$. Theorem~\ref{thm:NHC} then gives
\[
\|f_{\e}\|_2 ~\le~ \|f\|_{\kappa}, \quad \mbox{with} \quad \kappa ~=~ -\frac{\frac{Ent_q\(\frac{f}{\|f\|_1}\)}{n}}{\phi_{\e}\(1 - \frac{Ent_q\(\frac{f}{\|f\|_1}\)}{n}\)}.
\]

\noi Taking $q \rarrow 1$ and recalling that $Ent_q(\cdot) \rarrow_{q \rarrow 1} Ent(\cdot)$ completes the proof of the claim.

\noi We pass to the second claim of the corollary. First consider the case $\e = 0$. Note that in this case $q_0 = 2$. Furthermore, by the first claim of Lemma~\ref{lem:phi-derivatives}, $\phi_{0}(x) = \frac{x-1}{2}$, and hence the value of $\kappa$ given by the claim is $2$ (as expected).

\noi Assume now $\e > 0$. This implies that $q_0 < 2$, and hence, by the third claim of Lemma~\ref{lem:phi-technical}, for any $0 \le x \le 1$ we have $\frac{-x}{\phi_{\e}(1-x)} \le q_0 < 2 = q$. Hence the second clause in the definition of $\kappa_{2,q}$ in Definition~\ref{dfn:psi-kappa} does not apply. The remaining two clauses give the claim, as stated.

\eprf

\subsubsection*{Proofs of comments to Theorem~\ref{thm:NHC}}

\noi Some of the claims in these comments require a proof. These claims are restated and proved in the following lemma.

\lem
\label{lem:comm-thm}

\begin{itemize}

\item If $q \ge 2$ then for any $0 < \e < \frac12$ the function $\kappa_{2,q}(x,\e)$ starts as a constant-$\(1 + (1-2\e)^2\)$ function up to some $x = x(q,\e) > 0$, and becomes strictly decreasing after that. For $1 < q < 2$ there is a value $0 < \e(q) < \frac12$, such that for all $\e \le \e(q)$ the function $\kappa_{2,q}(x,\e)$ is strictly decreasing (in which case we say that $x(q,\e) = 0$). However, $x(q,\e) >  0$ for all $\e > \e(q)$. The function $\e(q)$ decreases with $q$ (in particular, $\e(q) = 0$ for $g \ge 2$). The function $x(q,\e)$ increases both in $q$ and in $\e$.

\item The function $\kappa_{2,1}(x,\e) = -\frac{x}{\phi_{\e}(1-x)}$ is strictly decreasing in its first argument for any $0 < \e < \frac12$. It satisfies $\kappa_{2,1}(0,\e) = \lim_{x \rarrow 0} \kappa_{2,1}(x,\e) = 1 + (1-2\e)^2$, for all $0 \le \e \le \frac12$.

\item Let $f$ be a non-constant function on $\H$. Let $0 < \e < \frac12$. Let $F(q) = F_{f,\e}(q) = \kappa_{2,q}\(Ent_q\(\frac{f}{\|f\|_1}\)/n,\e\)$. There is a unique value $1 < q(f,\e) \le 1 + (1-2\e)^2$ of $q$ for which $F(q) = q$. Moreover, $q(f,\e) = \min_{q \ge 1} F(q)$. Furthermore, $\lim_{\e \rarrow 0} q(f,\e) = 2$ for any $f$.

\end{itemize}

\elem

\prf
The first claim of the lemma follows from the properties of $\kappa_{2,q}$ as shown in the proof of Theorem~\ref{thm:NHC}. In particular, it is easy to see that for $q \le 2$ we have $\e(q) = \frac{1 - \sqrt{q-1}}{2}$ and for $\e \ge \e(q)$ we have $x(q,\e) = \frac{q - \(1+(1-2\e)^2\)}{\(1+(1-2\e)^2\) \cdot (q-1)}$. The claim that $\e(q)$ decreases with $q$ and that $x(q,\e)$ increases in both $q$ and $\e$ follows by direct verification.

\noi The second claim of the lemma follows immediately from the third claim of Lemma~\ref{lem:phi-technical}.

\noi We pass to the third claim of the lemma. Note that the function $x(q) = Ent_q\(\frac{f}{\|f\|_1}\)/n$ is positive and strictly increasing in $q$. We need the following auxiliary claim.

\lem
\label{lem:auxiliary}
The function $y(q) = \frac{q-1}{q} \cdot x(q) + \frac 1q$ is strictly decreasing in $q$.
\elem

\prf (of Lemma~\ref{lem:auxiliary})

\noi Assume w.l.o.g. that $f \ge 0$ and that $\E f = 1$. Let $P = \frac{f}{2^n}$ be a distribution on $\H$. A simple calculation gives that
\[
y(q) ~=~ 1 + \frac 1n \cdot \log_2 \(\(\sum_{a \in \H} P(a)^q\)^{\frac 1q}\),
\]
which is strictly decreasing in $q$, by H\"older's inequality.
\eprf

\noi We proceed with the proof of of the third claim of Lemma~\ref{lem:comm-thm}. Let $q_0 = 1 + (1-2\e)^2$. We claim, first, that $F$ is strictly increasing on $q_0 \le q < \infty$. In fact, for these values of $q$ the second clause of Definition~\ref{dfn:psi-kappa} does not apply (by the third claim of Lemma~\ref{lem:phi-technical}) and we have
\[
\kappa_{2,q}(x,\e) ~=~ \left\{\begin{array}{ccc} q_0 & \mbox{if} & y \le \frac{1}{q_0} \\
\frac{\alpha_0 - 1}{\phi_{\e}(\alpha_0)} & \mbox{if} & y  >  \frac{1}{q_0} \end{array}\right.,
\]
where $y = y(q)$ and $\alpha_0$ is determined by $1 - \alpha_0 - \frac{\alpha_0 \phi_{\e}(\alpha_0)}{1-\alpha_0} = y$. The claim now follows by combining Lemma~\ref{lem:auxiliary}, and the third and fourth claims of Lemma~\ref{lem:phi-technical}.

\noi Next, we claim that there exists a unique value $1 \le q = q^{\ast} \le q_0$ for which $\frac{-x}{\phi_{\e}(1 - x)} = q$ (here $x = x(q)$). Moreover, $F$ decreases for $1 \le q \le q^{\ast}$ and increases for $q \ge q^{\ast}$. Finally, $F\(q^{\ast}\) = q^{\ast}$. Observe that verifying these claims will essentially complete the proof of the third claim of Lemma~\ref{lem:comm-thm} (apart from the fact that $\lim_{\e \rarrow 0} q(f,\e) = 2$).

\noi In fact, by the first and third claims of Lemma~\ref{lem:phi-technical}, and the fact that $x$ is strictly increasing in $q$, the function $\frac{-x}{\phi_{\e}(1 - x)}$ is strictly decreasing in $q$, taking values between $\frac{2}{\log_2\(\frac{4}{q_0}\)}$ and $q_0$. This means that it has a unique intersection $q = q^{\ast}$ with the function $q$ in $[1, q_0]$. Next, observe that by Definition~\ref{dfn:psi-kappa}  for $q \le q_0$ we have
\[
\kappa_{2,q}(x,\e) ~=~ \left\{\begin{array}{ccc} -\frac{x}{\phi_{\e}(1 - x)} & \mbox{if} &  -\frac{x}{\phi_{\e}(1 - x)} \ge q \\
\frac{\alpha_0 - 1}{\phi_{\e}(\alpha_0)} & \mbox{if} &  -\frac{x}{\phi_{\e}(1 - x)} \le q \end{array}\right.
\]

\noi This means that for $q < q^{\ast}$ we have $F(q) = \kappa_{2,q}(x,\e) = -\frac{x}{\phi_{\e}(1 - x)}$, which is decreasing in $q$, and for for $q > q^{\ast}$ we have $F(q) = \frac{\alpha_0 - 1}{\phi_{\e}(\alpha_0)}$, which increases in $q$ . Finally, for $q = q^{\ast}$, we have $F(q) = -\frac{x}{\phi_{\e}(1 - x)} = q$.

\noi It remains to verify that $\lim_{\e \rarrow 0} q(f,\e) = 2$. By the first claim of Lemma~\ref{lem:phi-derivatives}, $\phi_{0}(x) = \frac{x-1}{2}$. This means that for any $0 < x \le 1$ we have $\lim_{\e \rarrow 0} \frac{-x}{\phi_{\e}(1 - x)} = 2$. The claim follows since, by the preceding discussion, $q = q(f,\e) = \frac{-x(q)}{\phi_{\e}(1 - x(q))}$.

\eprf

\section*{Acknowledgments}

\noi We would like to thank Or Ordentlich for a very helpful discussion.

\end{document}